\documentclass{aa}
\usepackage{graphicx}
\usepackage{txfonts}
\usepackage{subfigure}
\begin{document}

\title{Mirrors for X-ray telescopes: Fresnel diffraction-based computation of point spread functions from metrology}

\author{L. Raimondi
          \inst{1}
          \and
          D. Spiga\inst{2}}        
\institute{Sincrotrone Trieste ScpA, S.S. 14 km 163.5 in Area Science Park, 34149 Trieste (TS) - Italy\\ \email{lorenzo.raimondi@elettra.trieste.it}
               \and
               INAF/Osservatorio Astronomico di Brera, Via E. Bianchi 46, 23807 Merate (LC) - Italy \\ \email{daniele.spiga@brera.inaf.it}}
 \date{Received 3 Sep 2014 / Accepted 29 Sep 2014}
 
\abstract
{The imaging sharpness of an X-ray telescope is chiefly determined by the optical quality of its focusing optics, which in turn mostly depends on the shape accuracy and the surface finishing of the grazing-incidence X-ray mirrors that compose the optical modules. To ensure the imaging performance during the mirror manufacturing, a fundamental step is predicting the mirror point spread function (PSF) from the metrology of its surface. Traditionally, the PSF computation in X-rays is assumed to be different depending on whether the surface defects are classified as figure errors or roughness. This classical approach, however, requires setting a boundary between these two asymptotic regimes, which is not known a priori.}
{The aim of this work is to overcome this limit by providing analytical formulae that are valid at any light wavelength, for computing the PSF of an X-ray mirror shell from the measured longitudinal profiles and the roughness power spectral density (PSD), without distinguishing spectral ranges with different treatments.} 
{The method we adopted is based on the Huygens-Fresnel principle for computing the diffracted intensity from measured or modeled profiles. In particular, we have simplified the computation of the surface integral to only one dimension, owing to the grazing incidence that reduces the influence of the azimuthal errors by orders of magnitude. The method can be extended to optical systems with an arbitrary number of reflections -- in particular the Wolter-I, which is frequently used in X-ray astronomy -- and can be used in both near- and far-field approximation. Finally, it accounts simultaneously for profile, roughness, and aperture diffraction.}
{We describe the formalism with which one can self-consistently compute the PSF of grazing-incidence mirrors, and we show some PSF simulations including the UV band, where the aperture diffraction dominates the PSF, and hard X-rays where the X-ray scattering has a major impact on the PSF degradation. The results are validated with ray-tracing simulations, or by comparison with the analytical computation of the half-energy width based on the known scattering theory, where these approaches are applicable. Finally, we validate this by comparing the simulated PSF of a real Wolter-I mirror shell with the measured PSF in hard X-rays.}
{}

\titlerunning{Mirrors for X-ray telescopes: Fresnel diffraction-based}
\authorrunning{L. Raimondi \and D. Spiga}
\maketitle

 \section{Introduction}\label{intro}
Optics for imaging X-ray telescopes consist of a variable number of coaxial grazing-incidence, double-reflection X-ray mirrors. Most X-ray telescopes have so far adopted the Wolter-I profile, achieving a double reflection on a grazing-incidence paraboloidal mirror segment and a hyperboloidal one (Van Speybroeck \& Chase \cite{VanSpeyChase}): accurate on-axis focusing is obtained by means of two consecutive reflections onto these two surfaces. Alternative solutions can be envisaged, for instance, polynomial profiles (Conconi \& Campana~\cite{Conconi2001}; Conconi~et al.~\cite{Conconi2010}), to enlarge the optical field of view, or Kirkpatrick-Baez geometries (Kirkpatrick \& Baez~\cite{Kirkpatrick1948}), but all solutions rely on two or more reflections. In addition to the intrinsic aberrations of the optical design, especially off-axis, the mirror surface accuracy determines the concentration and the imaging performances. These quantities are typically expressed in X-ray astronomy using the point spread function (PSF), that is, the annular integral of the focused X-ray intensity around the center of the focal spot. Another quantity of frequent use to denote the imaging properties is the half-energy width (HEW), that is, twice the median value of the PSF. Achieving optical systems with high angular resolution -- for example, lower than 5 arcsec HEW for the ATHENA X-ray observatory (Willingale~et al.~\cite{Willingale2014}; Bavdaz~et al.~\cite{Bavdaz}) that is to be launched in 2028 -- requires accurate mirror metrology over a wide range of spatial scales, and also methods for predicting the PSF from metrology data at various X-ray energies. 

Mirror imperfections affecting the PSF in X-rays are traditionally divided into figure errors, for instance measured with optical profilometers (Tak\'{a}cs~et al.~\cite{Tacacs}), and microroughness, which can be measured with techniques like phase-shift interferometry (PSI, see, e.g., Upputuri~et al.~\cite{PSI}) or atomic-force microscopy (AFM, see, e.g., Dixson~et al.~\cite{AFM}). Except for definitions that empirically refer to the mirror length (De~Korte~et al.~\cite{DeKorte}), the separation of profile geometry and roughness in general reflects the different treatments adopted to predict their impact on the angular resolution. For example, if a measured profile is decomposed into Fourier components, profile errors encompass long spatial wavelengths, where geometrical optics can be applied. According to this definition, the PSF of a mirror characterized by defects of this kind can be predicted by ray-tracing routines that reconstruct the path of rays reflected at different mirror locations, regardless of the X-ray wavelength, $\lambda$. This method can be readily extended to multiple reflection systems, such as the Wolter system. In contrast, the surface roughness is assumed to entirely fall in a spectral region of spatial wavelengths where the concept of ``ray'' is no longer applicable because the optical path differences introduced by the roughness start to be similar to $\lambda$. In this spectral range, the PSF broadening stems from the wavefront diffraction off the reflecting surface, or X-ray scattering (XRS), which in general increases in intensity with the X-ray energy. The characterization of the microroughness over a wide range of spatial frequencies is conveniently expressed in terms of its power spectral density (PSD), because its values do not depend on the measurement technique in use (\cite{ISO10110}). Moreover, a well-established first-order theory can be very effectively used (Church~et al.~\cite{Church79}; Stover~\cite{Stover95}) to compute the scattering diagram, provided that the smooth surface condition is fulfilled,
\begin{equation}
	 4\pi \sigma\sin\alpha_0<\lambda,
	\label{eq:smooth}
\end{equation}
where $\alpha_0$ is the grazing (i.e., measured from the surface) incidence angle of X-rays and $\sigma$ is its surface error root mean square (rms) in a given spectral band. A noticeable result of this scattering theory is that the XRS angular distribution is simply proportional to the PSD. On this basis, several contributions (Christensen~et al.~\cite{Christensen1988}; Willingale~\cite{Willingale1988}; O'Dell~et al.~\cite{Odell1993}) were given in the past years to establish a relationship between the mirror PSF and the surface finishing level. One of us (Spiga \cite{Spiga2007}) has used the scattering theory to derive analytical formulae that can be used to convert the surface PSD of a mirror into the X-ray scattering term of the HEW as a function of $\lambda$ and vice versa. However, the first-order theory cannot be always extended to the low-frequency limit, where the surface defects are usually of larger amplitude. This limit has been overcome, for example, by Harvey~et al.~(\cite{Harvey1988}), who provided a transfer function-based approach to relate the PSF to the self-correlation function of a stochastically rough surface. More recently, a complete approach to the modeling of light scattering from rough surfaces has been provided by Schr\"oder~et~al.~(\cite{Schroeder}) to bridge the gap between scattering theories that are valid in different smoothness conditions.

\begin{figure}[!tpb]
\centering
    \includegraphics[width=0.5\textwidth]{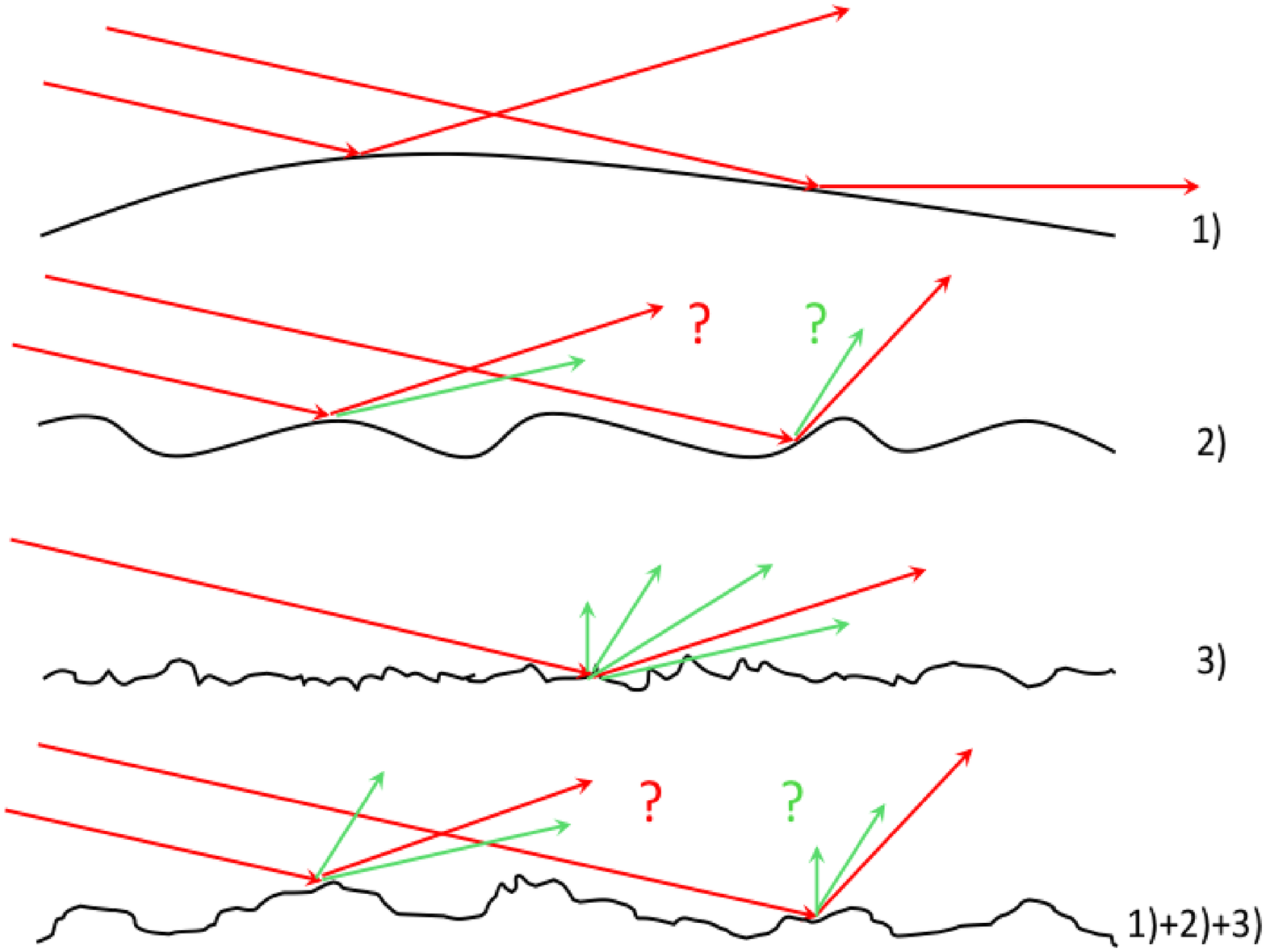}
    \caption{Different spatial wavelengths in a mirror profile error. Long wavelengths (1) are usually treated with geometrical optics, high-frequency roughness components (3) with the first-order scattering theory. The treatment of mid-frequencies (2) is more uncertain. Even more uncertain is the most general situation, in which all three components contribute to the mirror PSF.}
    \label{fig:mid-freq}
\end{figure}

The approaches just mentioned always require a spatial frequency that serves as a boundary between figure errors (non-stochastic) and roughness (stochastic); however, this limiting frequency is not of immediate definition. For this reason, adopting the geometric or scattering treatment has for a long time been ``a matter of taste'', to quote Aschenbach (\cite{Aschenbach2005}), who is credited to have shed some light on solving this problem on physical grounds. Aschenbach concluded that any single Fourier component whose $\sigma$ fulfills the smooth-surface condition (Eq.~(\ref{eq:smooth})) should be mostly treated as roughness, and as figure error otherwise. This approach highlighted that the geometry or scattering treatment is not fixed, but should at least depend on $\alpha_0$ and $\lambda$. However, there are some drawbacks in this statement:
\begin{itemize} 
	\item{The criterion operates a selection on the rms values of a spectrum of discrete frequencies, therefore it is difficult to apply to a continuous PSD since the ``single component'' rms, and consequently the boundary frequency, would depend on the spectral resolution of the metrological instrument in use.}
	\item{The separation between the two regimes is not abrupt in reality (Sect.~\ref{singrat}). For example, the spatial frequencies near the smooth-surface limit cannot be treated in either way (Fig.~\ref{fig:mid-freq}). We refer to these components, often found in the centimeter-millimeter range of spatial wavelengths, as mid-frequencies.}
	\item{Every spectral range, treated separately, returns a PSF. We then have as many PSFs as the number of spectral ranges in which we have decomposed the profile, which should now be combined into a single, total, predicted PSF. Even if a convolution of the PSFs might seem a natural approach, this is not correct in general, as we show in Sect.~\ref{farfield}.}
\end{itemize}

These points highlight the need for a self-consistent method to predict the PSF from a complete metrology dataset, including profile, mid-frequencies, and roughness. To this end, wavefront propagation methods (i.e., applications of the Huygens-Fresnel principle) should be used to treat surface defects of any frequency, at any wavelength of the incident radiation. The reason is that the validity of the Huygens-Fresnel principle is unrestricted. The geometrical optics results are automatically found in the limit of large surface defects, or $\lambda \rightarrow 0$. 

Methods based on physical optics are frequently used in normal-incidence mirrors for visible light (see, e.g., Cady~\cite{Cady12}). For grazing-incidence mirrors, they were mostly used to model the X-ray scattering when the smooth-surface and small scattering angle conditions are not met, for example, by Beckmann \& Spizzichino~(\cite{BeckSpizz}), and Zhao \& Van Speybroeck~(\cite{Zhao2003}): nevertheless, they seem to have restricted this method to solely compute the XRS. Mieremet \& Beijersbergen (\cite{MierBeijer}) used the Huygens-Fresnel principle to evaluate the impact of the aperture diffraction in silicon pore optics, but did not include mirror defects in their analysis. Others adopted the wavefront propagation to interpret the results of X-ray mirror tests in visible or ultraviolet light (e.g., Saha~et al.~\cite{Saha2010}), but the analysis was limited to the case of a focus at an infinite distance from the mirror, that is, to far-field conditions. In fact, owing to the long focal lengths at play in X-ray astronomy, the far-field condition is fulfilled in a number of cases. However, it is not applicable to the optical systems in which two or more reflections occur in sequence within a short distance, like Wolter-I profiles and most of polynomial configurations.

In this work we describe in detail a method for computing the PSF -- and consequently the HEW -- of grazing-incidence X-ray optical systems, including the Wolter system, from measured or modeled profiles, simply making use of the Fresnel diffraction theory. We have already anticipated some results in previous papers (Raimondi \& Spiga~\cite{RaiSpi2010}; Raimondi \& Spiga~\cite{RaiSpi2011}; Spiga \& Raimondi~\cite{SpiRai2014}). Here we provide a complete derivation of the results, and extend the formalism to anisotropic sources, or to sources located at a finite distance. Even if the Fresnel diffraction theory is often used to compute the PSF accounting for diffraction aperture and optical aberrations, it seems not to have been applied to real mirror profiles, that is, accounting for profile errors and roughness in a very wide spectral range of spatial frequencies. To this end, we need to reduce the surface integrals to only one dimension, corresponding to the longitudinal axis (Sect.~\ref{approx}). We provide simple formulae to compute the field diffracted by a grazing-incidence mirror profile at any light wavelength, with simplified expressions in the far-field case (Sect.~\ref{field}). In Sect.~\ref{wolter} we extend the formalism to double-reflection mirrors, which are often adopted in X-ray astronomy, and we show in Sect.~\ref{PSFcomp} some examples of PSF computations using these formulae. The geometrical optics results are automatically obtained at X-ray energies at which aperture diffraction and X-ray scattering are negligible. In certain conditions, we can even compare the HEW($\lambda$) values computed from Fresnel diffraction with the results obtained from the analytical treatment (Spiga~\cite{Spiga2007}) of the XRS term of the HEW. A very good agreement is found between the two methods (Sect.~\ref{HEW}), provided that the XRS term and the figure error term of the HEW are summed linearly. We also report in Sect.~\ref{exper} an experimental verification of the predictions for a hard X-ray mirror shell tested at the SPring-8 radiation facility (Spiga et al.~\cite{Spiga2011}). A short summary of the results is given in Sect.~\ref{concl}.
\begin{figure}[!tpb]
\centering
   \resizebox{\hsize}{!}{\includegraphics{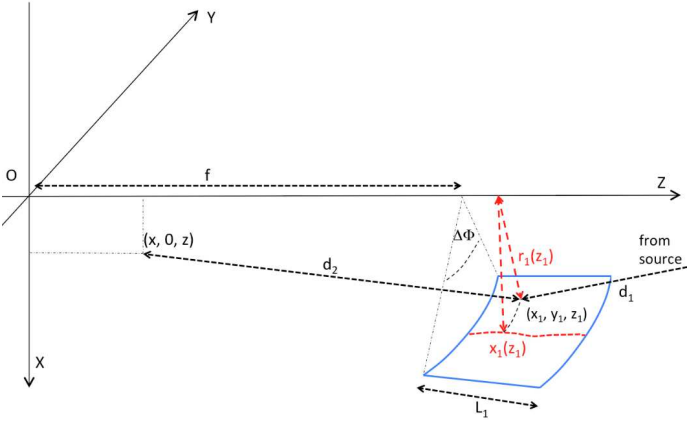}}
    \caption{Reference frame used to compute the diffracted field from a grazing-incidence mirror. The scattered amplitude at the generic point in the $xz$ plane is obtained by superposing secondary waves generated at each point of the mirror profile ($x_1$, $y_1$, $z_1$).}
    \label{fig:refframe}
\end{figure}

\section{Grazing incidence and monodimensional approximation}\label{approx}
Wavefront propagation techniques are widespread in optics to assess the impact of the aperture diffraction effects on the imaging quality. Indeed, this method has rarely been applied to real mirrors with measured surface defects. The reason is that most codes for wavefront propagation are two-dimensional, meaning that they compute the intensity distribution over a 2D focal plane from a 2D surface mirror map. This makes the computation quite intensive, however, therefore it can only be applied to profiles that are known analytically or to profiles whose measured shape is sampled with a convenient lateral step ($\ge$~1~mm). This clearly rules out including the roughness in the PSF computation because this would imply a sampling step typically below 1~$\mu$m, and so the number of iterations required would be larger by more than a factor of $10^6$! In contrast, the computation is enormously simplified when the Huygens-Fresnel principle is applied to 1D profiles in the axial direction. In addition, this enables us to compute the PSF on a single line in the focal plane, averaging the results if several axial profiles have to be analyzed.

The method we hereby provide is exactly based on a 1D computation and can be applied to a variety of cases. For example, the astronomical case, with a source at a practically infinite distance from a mirror focusing via a double-reflection at a shallow angle. X-ray mirrors or mirror assemblies of this kind are also tested using terrestrial sources such as MPE/PANTER (Burwitz~et al.~\cite{Burwitz}), where a very small X-ray source is located at a finite, although very large, distance. Among other effects (Van Speybroeck \& Chase \cite{VanSpeyChase}), the finiteness of the source distance causes a small, intrinsic defocusing in Wolter-I mirrors, but this is in general negligible with respect to the influence of fabrication errors. X-ray mirrors are also used at terrestrial X-ray sources like synchrotron radiation facilities or free electron lasers (FELs) such as FERMI at Elettra (Allaria~et al.~\cite{Allaria}), where an X-ray beam of noticeable spatial (also temporal in FELs) coherence is generated from a very small source. In these cases, the high source brilliance does not require a tight mirror nesting; higher focusing performances are usually required, and because of the finiteness of the source distance, an exact focusing in single reflection can be obtained only using ellipsoidal mirrors. If the mirror is characterized by a very high profile accuracy and surface finishing, then the source size and its coherence properties have also to be taken into account (Raimondi~et al.~\cite{RaimondiNIMA2013}).

We consider throughout a radiation of wavelength $\lambda$, propagating in the negative $z$ direction of the reference frame (see Fig. \ref{fig:refframe}) and impinging on an axially-symmetric grazing-incidence mirror, of length $L_1$ and optical axis coincident with the $z$-axis. The mirror is a sector with an azimuthal aperture $\Delta \Phi$, with linear dimensions much larger than $\lambda$ to avoid azimuthal diffraction effects. In the axial direction, the mirror spans from $f$ to $f+L_1$, and the azimuthal (sagittal) radius of the mirror also increases. We denote the theoretical radius at $z = f$ with $R_0$ and the radius at the other end with $R_{\mathrm M}$. 

The wavefront is assumed to be initially uniform and spherical, with an electric field amplitude $E_0$ at $z = f+L_1$. The source is a point located at $z = S$. The wavefront can initially diverge ($S \gg0$) or converge ($S \ll 0$), but we always assume $|S| \gg L_1$. The axial profile of the mirror, including defects resulting from profile, mid-frequencies and roughness, is described by the coordinate array ($x_1$, $z_1$) in the $xz$ plane. For simplicity, we assume the mirror system to focus the radiation from the source to the $z =0$ plane at a distance $f$ out to the mirror's nearest end. We explicitly point out that $f$ is the mirror distance needed to have the best focal plane at $z =0$, which in general differs from the mirror focal length unless the source is located at infinity. For simplicity, we also assume $|S| > f$ and define $D = S-f$ as source-to-mirror distance.

We now define $\alpha_0$ to be the incidence angle at $z = f$ for a source at infinite distance, measured from the surface: $\alpha_0$ must be shallow (smaller than a few degrees), otherwise the reflectivity will be very low. The variation of the incidence angle on the mirror in a meridional plane is in general even smaller than $\alpha_0$ itself (Spiga~et al.~\cite{Spiga2009}), even though some curvature is obviously needed for the mirror to have a focus. So we may write, to a good approximation, that $R_{\mathrm M}-R_0 \simeq L_1\sin\alpha_0$. In the general case we also have to account for the divergence, or the convergence, of the incoming beam. The divergence angle of the wave -- also approximately constant -- is denoted with $\delta = R_0/D$, taken with the same sign of $D$: we can therefore write the incidence angle on the mirror as
\begin{equation}
	\alpha_1 = \alpha_0+\delta,
	\label{eq:alpha1}
\end{equation}
and the radial amplitude of the mirror's entrance pupil seen from the source as
\begin{equation}
	\Delta R_1 \simeq L_1\sin\alpha_1.
	\label{eq:rad_ampl}
\end{equation}

\begin{figure}[!tpb]
\centering
    \includegraphics[width=0.5\textwidth]{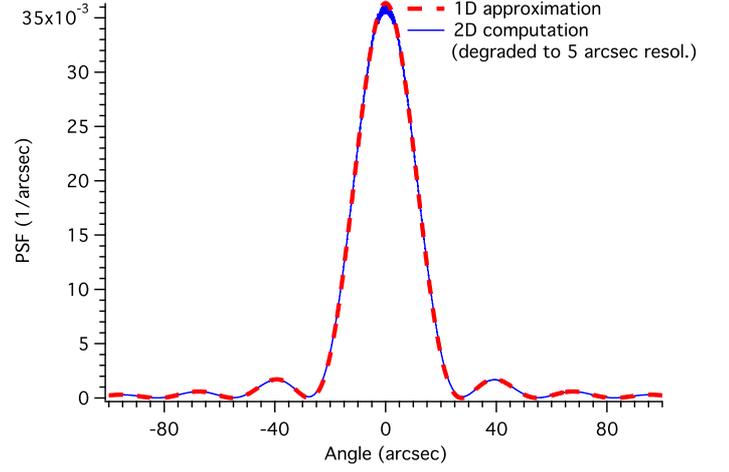}
    \caption{Computed aperture diffraction PSF at $\lambda$ = 3000~\AA, source at infinity, of a grazing-incidence parabolic mirror with $f$ = 10~m, a minimum radius $R_0$ = 150~mm, and a length $L_1$ = 300~mm, resulting in a circular corona aperture of 2.25~mm width. The dashed line is the usual diffraction pattern of a straight slit of equal width, while the accurate computation (solid line) is obtained by computing the exact diffraction pattern integrated over circular coronae. A high-frequency modulation in the latter would also be superimposed owing to the diffraction of the $2R_0$ diameter circular aperture, but in real cases is canceled out by the finite resolution of the detector.}
    \label{fig:diffpat}
\end{figure}

Our aim is to devise a general formula for the mirror PSF, defined as the diffracted field intensity at $z =0$, integrated in circular coronae between the radii $x$ and $x+\Delta x$, divided by $\Delta x$, and normalized to the intensity collected by the mirror. Doing this, assuming shallow incidence angles has two advantages: 
\begin{enumerate}
	\item{the two polarization states do not practically change direction after reflection, so we can readily work in scalar approximation;}
	\item{we can limit the computation to the longitudinal (i.e., axial) profiles, neglecting thereby the transverse deflection caused by azimuthal (i.e., sagittal) errors.}
\end{enumerate}
The last point, which allows reducing the computation complexity dramatically, is justified by the following considerations:
\begin{itemize}
	\item{If geometrical optics can be applied, the slope errors of the longitudinal sections of the mirror result in an angular dispersion twice as large, while the same slope errors along the azimuth result in an angular spread of rays smaller by a factor of $\tan2\alpha_1$.}
	\item{The scattering in the incidence plane, determined by the roughness PSD computed in the longitudinal direction, is more extended (Church~\cite{Church88}) than in the perpendicular direction by a factor of $(\tan\alpha_1)^{-1/2}$; in other words, the XRS pattern is almost unaffected by the profiles along the azimuth.}	
	\item{Since the mirror aperture is a circular corona of width $\Delta R_1~\ll~R_0$, also the aperture diffraction -- visible when the mirror is tested in UV light -- resembles the diffraction pattern of a long, straight slit, which can be computed in 1D (an example is shown in Fig.~\ref{fig:diffpat}).}
\end{itemize}

An implication of the 1D approximation is that the PSF abruptly drops just out of the incidence plane; hence, at any azimuthal coordinate of the mirror the PSF collapses into a single line, and the intensity distribution along the line is a function only of the radial distance from the center of the focal spot. Then the integration in circular coronae is made immediately, and the PSF becomes a function of the sole coordinate $x$. In this way, it is sufficient to compute the PSF along the $x$-axis instead of throughout the entire detector area.

We thereby assume the mirror surface to be described as a rotation of a 1D profile about the $z$-axis, that is the radial coordinate as a function of the mirror's axial coordinate, $r = r_1(z_1)$, which in turn equals the longitudinal mirror profile in the $xz$ plane $x_1(z_1)$. In practice, $x_1$ is composed of three terms:
\begin{equation}
	x_1(z_1) = x_{{\mathrm n}1}(z_1)+x_{\mathrm{meas}1}(z_1)+x_{\mathrm{PSD}1}(z_1),
	\label{eq:profile_comp}
\end{equation}
where $x_{{\mathrm n}1}$ is the nominal mirror profile, $x_{\mathrm{meas}1}$ is the measured profile error along the entire profile length $L_1$, and $x_{\mathrm{PSD}1}$ is one of the infinitely possible profiles of length $L_1$, computed from the PSD. The latter is to be obtained from a previous roughness characterization in a broad spectral range, but not overlapping the frequency window of the instrument used to measure $x_{\mathrm{meas}1}$. The reason for the different treatment for the two terms is that the resolution of $x_{\mathrm{meas}1}$ cannot be extended down to the typical frequencies of microroughness. Conversely, instruments dedicated to roughness measurements cannot be extended to scan lengths of more than a few millimeters. Hence, the PSD characterization can be used to obtain one of the infinitely possible profiles of length $L_1$ (Sect.~\ref{rough}) that are consistent with the measured roughness PSD. The reason for the profile degeneracy lies in the phase information of the Fourier components of the roughness, which are lost when computing the PSD. To reconstruct the profile from the PSD, the phase of the components can be freely selected. Each choice results in a different rough profile, which in principle might exhibit different scattering properties. 

Fortunately, one of the results of the first-order XRS theory is that the scattering pattern only depends on the PSD if the rms of $x_{\mathrm{PSD}1}$ fulfills Eq.~(\ref{eq:smooth}):
\begin{equation}
	\left(\int_0^{L_1}\! x_{\mathrm{PSD}1}\,\mbox{d}z_1\right)^{1/2}\! < \frac{\lambda}{4\pi\sin\alpha_1}.
	\label{eq:roughlim}
\end{equation}
Equation~(\ref{eq:roughlim}) is usually fulfilled by optically polished surfaces, therefore we expect the PSF contribution of $ x_{\mathrm{PSD}1}$ to depend not on the particular realization of the rough profile, but on the sole PSD. 

We finally point out that the decomposition of $x_1(z_1)$ is purely operational, meaning that it is only related to the sensitivity of measurement methods used for different windows of spatial frequencies, and the condition of Eq.~(\ref{eq:roughlim}) is requested to $x_{\mathrm{PSD}1}$ only to reconstruct the profile reliably. We show below that the same formulae for the PSF can be applied, regardless of whether the smooth-surface condition is fulfilled or not.

\section{PSF of a grazing-incidence single-reflection mirror}\label{field}
\subsection{Isotropic, point-like source}\label{isopoint}
We consider the case of a point-like and isotropic source on the optical axis of a axially-symmetric, grazing-incidence mirror sector characterized by the radial profile $x_1(z_1)$, as described in Sect.~\ref{approx}. Referring to the scheme depicted in Fig.~\ref{fig:refframe}, the electric field diffracted in the $xz$ plane can be easily computed by means of the Huygens-Fresnel principle. The derivation, reported in Appendix~\ref{derivation}, returns the following expression (Eq.~(\ref{eq:HF_fin})):
\begin{equation}
	E(x,0,z)=\frac{E_0\, \Delta R_1}{L_1\sqrt{\lambda x}}\int_f^{f+L_1}\!\!\!\!\sqrt{\frac{x_1}{\bar{d}_2}}\,e^{-\frac{2\pi\mathrm{i}}{\lambda} \left[\bar{d}_2-z_1+\frac{x_1^2}{2(S-z_1)}\right]} \,\mbox{d}z_1,
	\label{eq:field}
\end{equation}
where $\Delta R_1$ is given by Eq.~(\ref{eq:rad_ampl}), we have omitted unessential phase factors, evaluated the radial coordinate at $x_1(z_1)$, and defined 
\begin{equation}
	\bar{d}_2 = \sqrt{(x_1-x)^2+ (z_1-z)^2}.
	\label{eq:d2_av_def}
\end{equation}
If the diffracted field does not encounter subsequent mirrors, this expression can be used to derive the mirror PSF at the nominal focal plane ($z = 0$). The diffracted intensity on the $x$-axis is
\begin{equation}
	I(x)=\frac{E^2_0\, (\Delta R_1)^2}{L_1^2\lambda x}\left|\int_f^{f+L_1}\!\!\!\!\sqrt{\frac{x_1}{\bar{d}_{2,0}}}\,e^{-\frac{2\pi \mathrm{i}}{\lambda} \left[\bar{d}_{2,0}-z_1+\frac{x_1^2}{2(S-z_1)}\right]} \,\mbox{d}z_1\right|^2,
	\label{eq:intensity}
\end{equation}
where $\bar{d}_{2,0}$ is Eq.~(\ref{eq:d2_av_def}) evaluated at $z = 0$. Owing to the symmetry about the $z$-axis, and since by hypothesis the sector is wide enough to avoid edge diffraction at it sides, Eq.~(\ref{eq:intensity}) is valid on the focal plane for azimuthal angles within [$-\Delta\Phi/2$, $+\Delta\Phi/2$]. Therefore, integrating the intensity on the focal plane over a circular segment of area $x\,\Delta\Phi\,\Delta x$, dividing by $\Delta x$, and normalizing to the intensity collected by the mirror slice $E_0^2 R_0\, \Delta\Phi\,\Delta R_1$, we obtain the formula for the PSF of a single-reflection grazing-incidence mirror:
\begin{equation}
	\mbox{PSF}(x)=\frac{\Delta R_1}{L_1^2\lambda R_0 }\left|\int_f^{f+L_1}\!\!\!\!\sqrt{\frac{x_1}{\bar{d}_{2,0}}}\,e^{-\frac{2\pi \mathrm{i}}{\lambda} \left[\bar{d}_{2,0}-z_1+\frac{x_1^2}{2(S-z_1)}\right]} \,\mbox{d}z_1\right|^2.
	\label{eq:PSF}
\end{equation}
If all the lengths in Eq.~(\ref{eq:PSF}) are measured in millimeters, the PSF is measured in mm$^{-1}$. To have the focal line graded in arcseconds, it is sufficient to multiply the $x$-axis times the plate-scale factor 206\,265$/f$ and divide the PSF by the same factor to have it measured in arcsec$^{-1}$. In that case, we denote the angular distance from the PSF center with $\theta$. We also note that
\begin{enumerate}
	\item{The derivation of Eq.~(\ref{eq:PSF}) is based on the Huygens-Fresnel principle and the grazing incidence approximation; therefore it is valid for any value of $\lambda$.}
 	\item{Numerical computation shows that the PSF is normalized to 1 if integrated over the entire $x$-axis (we prove this analytically, for a particular case, in Sect.~\ref{farfield}).} 
	\item{If the computation is performed over a focal line of finite size $2\rho$ (from now on called ``detector''), then the PSF integral is less than 1, because some beam is scattered out of the detector size. However, if the HEW is computed with respect to the absolute normalization, then its value is independent of $\rho$, on condition that the detector is wide enough for the PSF integral to exceed 1/2.}
\end{enumerate}
 
The PSF in Eq.~(\ref{eq:PSF}) is entirely determined by the function $x_1 (z_1)$: the real, longitudinal profile of the mirror, including its real defects, regardless of any distinction between figure errors, mid-frequencies, or roughness. If the profile is known analytically, then the integral can be explicitly solved, but this can only be done in a few cases. In general, the PSF is computed from a tabulated profile, with a finite spatial resolution, $\Delta z_1$, which has to be low enough to sample the shortest measured wavelength in the profile. However, it also needs to be short enough to avoid ghost features. The maximum sampling step of the profile is the spatial wavelength, $\lambda f/(\sin\alpha_1\,\rho)$, which causes a first-order scattering at the detector edge $x=\pm\rho$, halved to fulfill the Nyquist criterion and oversampled by a factor of $2\pi$ (Raimondi \& Spiga~\cite{RaiSpi2010}):
\begin{equation}
	\Delta z_1 = \frac{\lambda f}{4\pi \sin\alpha_1\,\rho}.
	\label{eq:minsamp_z}
\end{equation}
This sampling enables computing the PSF within the detector size. The measured profile error, $x_{\mathrm{meas}1}$, has to be at least sampled at this spatial step, and the PSD on the corresponding spatial frequencies [$1/L_1$, $2/L_1$, \ldots, $1/2\Delta z_1$], as per the Nyquist theorem. In turn, higher spatial frequencies (typically obtained from a roughness PSD measurement in the AFM range) may need to be included as well, up to a highest value $\nu_{\mathrm{max}}>(2\Delta z_1)^{-1}$: to this end, the sampling step in $z_1$ should be clearly reduced to $(2\nu_{\mathrm{max}})^{-1}$. Expanding the frequency band in the profile clearly increases the scattering amount out of the detector edge, which is compensated by a reduced PSF normalization. We note that Eq.~(\ref{eq:minsamp_z}) was derived from the grating formula at the first order of interference, but it remains valid at higher orders: in fact, the $2k\Delta z_1$ (for integer $k$) wavelength also contributes to a $k^{th}$ order scattering at the same angle, but this wavelength is well oversampled by step $\Delta z_1$ provided by Eq.~(\ref{eq:minsamp_z}). 

We can also derive the sampling requested for the detector, $\Delta x$, defined as the coordinate at which the minimum spatial frequency in the profile, $1/L_1$, scatters at the first order, oversampled by $2\pi$:
\begin{equation}
	\Delta x = \frac{\lambda f}{2\pi \sin\alpha_1\,L_1}.
	\label{eq:minsamp_x}
\end{equation}
The resulting number of sampled points, $N$, is the same for the mirror and for the detector:
\begin{equation}
	N = \frac{2\rho}{\Delta x} = \frac{L_1}{\Delta z_1} = \frac{4\pi \,\Delta R_1\, \rho}{\lambda f}.
	\label{eq:numsampl}
\end{equation}

The results just listed -- and in the remainder of this paper -- can be generalized to mirrors with non axially symmetric errors. If different sectors of a grazing-incidence mirror are characterized by different measured axial profiles, the PSF of each sector can be computed from the individual profiles, and the PSFs obtained can be averaged to return the final PSF. The profiles can also include tilt or offset errors with respect to the nominal profile of the mirror. The extension of the computation to a source off-axis in the $xz$ plane is straightforward, changing the definition of $d_1$ by Eq.~(\ref{eq:d1_spherical_off}) or~(\ref{eq:d1_planar_off}), provided that the off-axis angle $\theta_{\mathrm s}$ of the source is much smaller than $\alpha_1$.

The previous results are exactly valid only for a source of ideal temporal coherence, meaning a perfectly monochromatic source. To account for the finite coherence length $\Delta s_{\mathrm{coh}}$, one can apply Eq.~(\ref{eq:PSF}) by varying $\lambda$ at random with $x$ within a wavelength bandwidth $\Delta \lambda \simeq \lambda^2/(2\pi \Delta s_{\mathrm{coh}})$. This has the effect of smoothing out fine PSF features, which would be visible only with perfectly monochromatic radiation.

\subsection{An example: the sinusoidal profile error}\label{singrat}
\begin{figure*}[!htpb]
    \centering
    \subfigure[]{\includegraphics[width=0.45\textwidth]{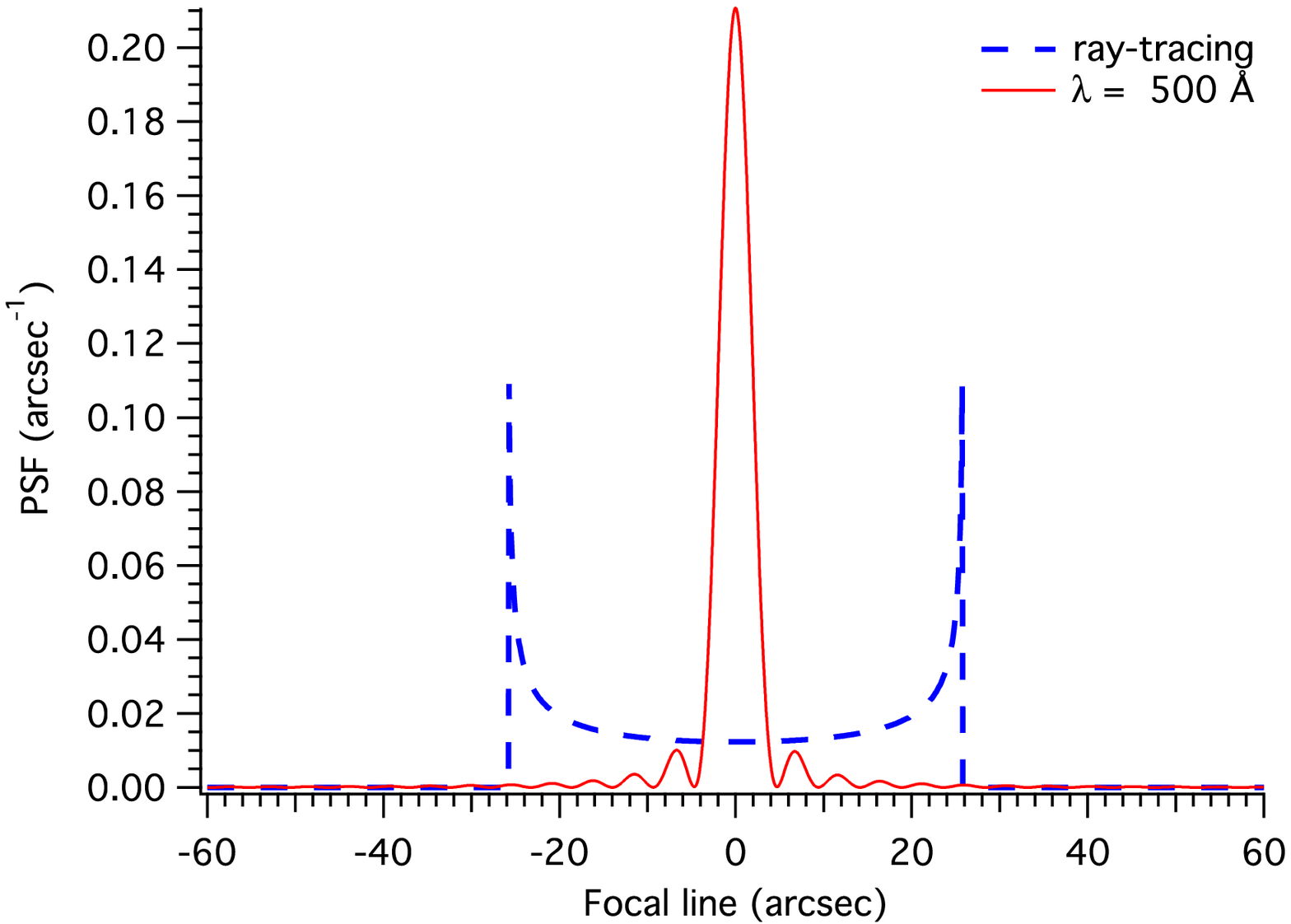}\label{fig:singrat_PSF_a}}
    \subfigure[]{\includegraphics[width=0.45\textwidth]{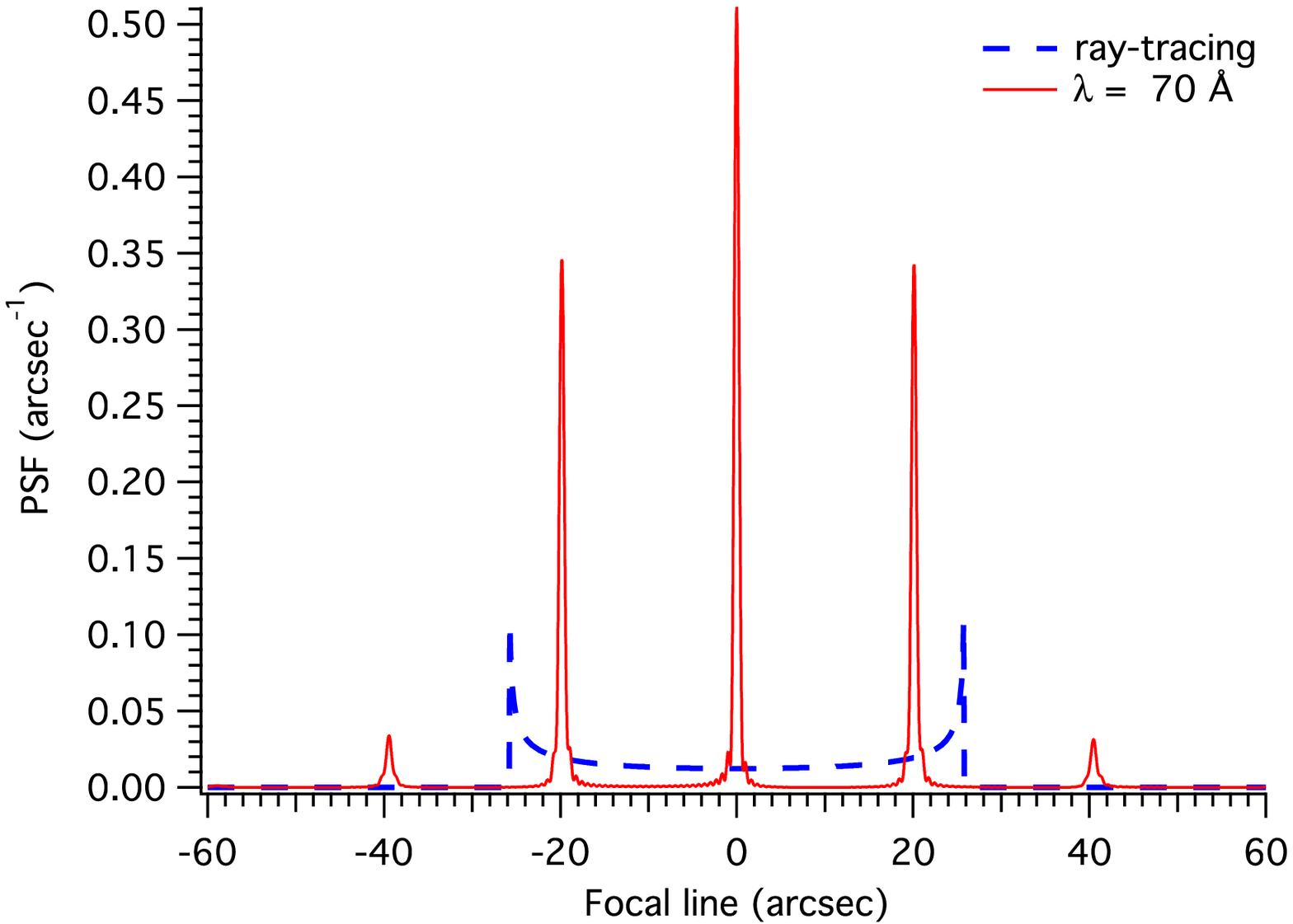}\label{fig:singrat_PSF_b}}
    \subfigure[]{\includegraphics[width=0.45\textwidth]{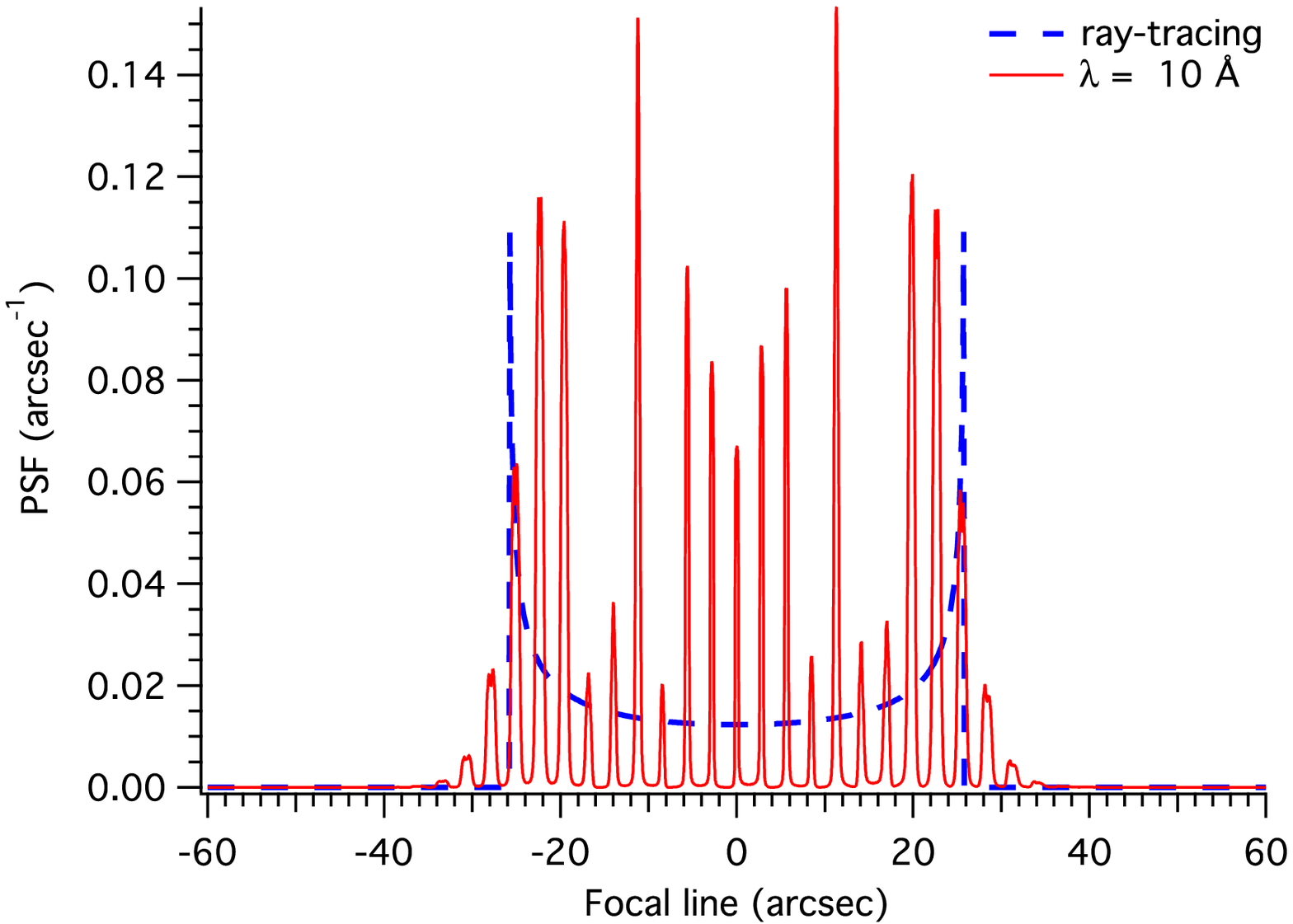}\label{fig:singrat_PSF_c}}
    \subfigure[]{\includegraphics[width=0.45\textwidth]{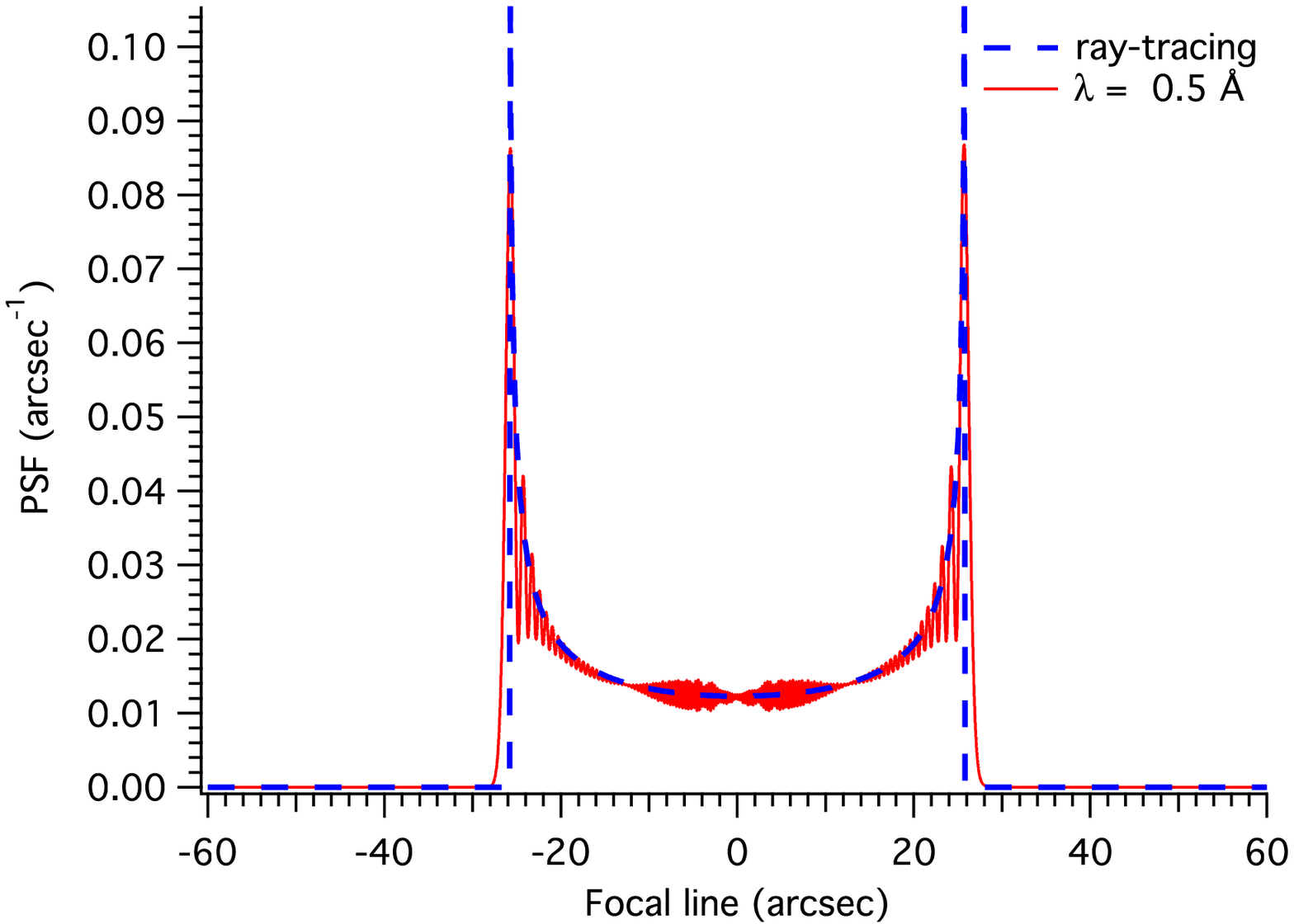}\label{fig:singrat_PSF_d}}
    \caption{Dashed lines: PSF expected from applying geometrical optics to a parabolic profile with a sinusoidal profile error with $A$~= 0.1~$\mu$m and $T$~= 10~mm (Eq.~(\ref{eq:singrat_PSF})). The parabolic profile parameters take on reasonable values $R_0$~= 15~cm, $f$~= 10~m, and $L_1$~= 300~mm. The detector has a resolution of 20 $\mu$m. Solid lines: computed PSF for decreasing $\lambda$, using Eq.~(\ref{eq:PSF}). (a) $\lambda$~= 500~\AA, dominated by aperture diffraction. (b) $\lambda$~= 70~\AA, first-order scattering dominates and the second-order peaks start to appear. (c) $\lambda$~= 10~\AA: multiple orders have become visible. (d) $\lambda$~= 0.5~\AA: high-order peaks are almost completely blended, and the PSF now resembles the geometrical optics result.}
    \label{fig:singrat_PSF}
\end{figure*}
In this section we show some applications of Eq.~(\ref{eq:PSF}) to the case of a parabolic nominal profile with a superposed sinusoidal pattern (Eq.~(\ref{eq:profile_comp})),
\begin{equation}
	x_{\mathrm{meas}1} = A \sin\left(\frac{2\pi}{T}z_1\right),
	\label{eq:singrat}
\end{equation}
and the source is assumed to be at infinity, therefore $\alpha_1 = \alpha_0$. Classically, if $T$ is in the centimeter range and the incidence angle is in the typical range of X-ray optics ($\approx$ 0.5~deg), this perturbation is difficult to classify as figure error or roughness, and accordingly falls in a mid-frequency range of uncertain treatment (Fig.~\ref{fig:mid-freq}). For example, if geometrical optics could be applied, the PSF would exhibit a typical diverging shape (Spiga~et al.~\cite{Spiga2013}):
\begin{equation}
	\mbox{PSF}(\theta) = \frac{1}{\pi} \left[\left(\frac{4\pi A}{T}\right)^2-\theta^2\right]^{-1/2}
	\label{eq:singrat_PSF}
\end{equation}
for $|\theta|< 4\pi A/T$, and zero elsewhere (dashed lines in Fig.~\ref{fig:singrat_PSF}). But in which conditions are we allowed to treat this sinusoidal perturbation with geometrical optics?

The application of Fresnel diffraction (Eq.~(\ref{eq:PSF})) allows us to overcome these uncertainties, and the results exhibit a more complicated picture. When $\lambda$ is in the UV range, the interferential pattern of the grating is invisible (Fig.~\ref{fig:singrat_PSF_a}), because it is completely hidden by the aperture diffraction. As $\lambda$ is diminished, the aperture diffraction decreases in proportion, and the PSF starts to resemble a Dirac delta, as it would for a perfect mirror. However, at sufficiently high energies, scattering peaks start to appear at the two sides of the central peak: at $\lambda$ = 70~\AA~the first-order peaks are the most prominent feature, while the second-order peaks appear (Fig.~\ref{fig:singrat_PSF_b}).

When the energy is increased (Fig.~\ref{fig:singrat_PSF_c}), the PSF becomes more complicated as peaks appear near the angles $\theta_k$: these angles are defined by the known grating equation
\begin{equation}
	T[\cos\alpha_0- \cos(\alpha_0+\theta_k)] = k\lambda,
	\label{eq:grat1}
\end{equation}
with $k$ integer. The peak height decays rapidly just beyond the angular range of the geometric PSF: the reason is that, as we show in Sect.~\ref{farfield}, in far-field and small scattering angle approximations the peak heights are 
\begin{equation}
	\mbox{PSF}(\theta_k) \propto J_k^2\left(\frac{4\pi A}{\lambda}\sin\alpha_0\right)
	\label{eq:grat2}	
\end{equation}
and $J_k$ is the $k^{th}$ Bessel function of the first kind. This is a known result of the sinusoidal grating theory. Now, for high values of $k>0$ and $|x| < k$, $J_k(x) \approx 0$. Hence, the PSF is nearly zero if 
\begin{equation}
	4\pi A \sin\alpha_0 < k\lambda
	\label{eq:grat3}
\end{equation}
for $k >0$. By comparison with Eq.~(\ref{eq:grat1}), we obtain
\begin{equation}
	4\pi A \sin\alpha_0 < T\,[\cos\alpha_0-\cos(\alpha_0+\theta_k)] 
	\label{eq:grat4}
\end{equation}
which no longer depends on $\lambda$. Developing Eq.~(\ref{eq:grat4}) yields
\begin{equation}
	\sin\frac{\theta_k}{2} > \frac{2\pi A}{T},  
	\label{eq:grat5}
\end{equation}
and a similar result is obtained for $k < 0$. Therefore, in the limit of small scattering angles, the PSF is near zero if 
\begin{equation}
	|\theta_k| > \frac{4\pi A}{T},  
	\label{eq:grat6}
\end{equation}
exactly like the result of geometrical optics. 

For very low values of $\lambda$ (Fig.~\ref{fig:singrat_PSF_d}), the separation between adjacent peaks eventually becomes smaller than the detector resolution and the peaks merge, forming a nearly continuous function that perfectly matches the PSF predicted by geometrical optics. Even with an ideal detector with infinite spatial resolution the peaks would merge, in practice because the peak spacing would be smoothed out by the finite monochromaticity of the X-ray source. This example shows that what we call ``geometric optics'' is nothing but the superposition of high scattering orders that blend for sufficiently low values of $\lambda$, and consequently, it can be simulated accurately using Eq.~(\ref{eq:PSF}), exactly like all the other physical optics effects!

We now return to the question for which conditions geometrical optics -- and consequently, ray-tracing programs -- can be applied. In general, there is no answer a priori. For example, the smooth-surface criterion (Eq.~(\ref{eq:smooth})) with $\alpha_0$ = 0.43~deg and $\sigma = A/\sqrt{2}$ = 71~nm is fulfilled for $\lambda > 67~\AA$. In fact, at $\lambda = 70~\AA$ the PSF is correctly dominated by the first-order scattering (Fig.~\ref{fig:singrat_PSF_b}), but the transition to geometrical optics is very gradual as the energy is increased: at $\lambda~= 10~\AA$ the computed PSF is still far from the geometrical optics predictions, and only for $\lambda < 1~\AA$ the ray-tracing results merge with the computation \`{a} la Fresnel (Fig.~\ref{fig:singrat_PSF_d}). 

A simple argument shows why the passage to the geometrical optics occurs near $\lambda =$ 1~\AA. Consider a profile patch of length equal to the spatial period $T$; the width seen by the X-ray beam is therefore $T\sin\alpha_0$, and the corresponding diffraction figure size at a distance $f$ is $2f\lambda/(T\sin\alpha_0)$. If the latter exceeds $T\sin\alpha_0$, that is,
\begin{equation}
	\lambda > \frac{T^2\sin^2\alpha_0}{2f}
	\label{eq:difflim}
\end{equation}
the size and the relief of the profile spatial period is completely hidden by the diffraction, i.e., the geometrical optics cannot be applied. Substituting the values one obtains $\lambda >$ 2.8~\AA, in good accord with the limit found via the simulation reported in Fig.~\ref{fig:singrat_PSF_d}. Other examples that show how the PSF reduces to the predictions of geometrical optics in the limit of small $\lambda$ or long spatial wavelengths are reported in the Sect.~\ref{PSFcomp} for double-reflection optical systems. 

\subsection{Extended and anisotropic sources} \label{incoh}
The results listed in the previous section are valid if the radiation source can be approximated by a geometric point. In these conditions the source is spatially coherent and isotropic, meaning that the wavefront is spherical and the electric field amplitude is the same in all the directions. For example, the X-ray source at PANTER (Burwitz~et al.~\cite{Burwitz}) has a 1~mm size out to a 123.9~m distance, so the point-like approximation is widely applicable as long as the mirror HEW is not better than the angular size of the source ($\sim$1.5~arcsec).  

\begin{figure}[!tpb]
     \centering
     \includegraphics[width=0.45\textwidth]{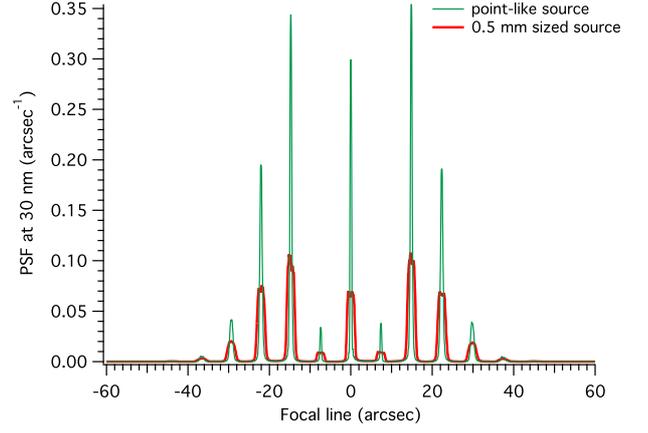}
     \caption{PSF simulation obtained from Eq.~(\ref{eq:PSF}) with an elliptical mirror, illuminated by a spatially incoherent source of 0.5 mm diameter at $\lambda =$ 30~\AA, with top-hat intensity profile. The parameters $f$, $R_0$, and $L_1$ are the same as for the simulations of Fig.~\ref{fig:singrat_PSF}. The radiation source is at a distance of 50~m from the mirror entrance. The peaks appear smoothed to the angular diameter of the source (2 arcsec). \label{fig:incoherent}}
\end{figure}

However, most astronomical sources, or X-ray facilities on the ground if the mirror PSF starts to become similar to the de-magnified source size (Raimondi~et al.~\cite{RaimondiSPIE2013}), are better represented with a finite extension. Most natural extended sources are spatially incoherent, and the image at the focal plane is obtained by decomposing the source into point-like sources of angular diameter $\phi_{\mathrm S}$ (Hol\'{y}~et al.~\cite{Holy}),
\begin{equation}
	\phi_{\mathrm S} < \frac{\lambda}{L_1\sin\alpha_1}
	\label{eq:coherent}
\end{equation}
at off-axis positions and intensities properly distributed within the source extent, using Eq.~(\ref{eq:PSF}), and superposing the diffraction patterns on the focal plane. The final result is a convolution of Eq.~(\ref{eq:PSF}) with the de-magnified intensity profile of the radiation source (Fig.~\ref{fig:incoherent}).

Other sources, such as synchrotrons and FELs, exhibit a high degree of spatial coherence and are markedly anisotropic; hence, the mirror illumination is often nonuniform, which clearly affects the measured PSF. For example, the FERMI at Elettra FEL1 is a coherent source with a Gaussian intensity profile (Svetina~et al.~\cite{Svetina2013}). The subsequent propagation of the wavefront stems from the source self-diffraction and, because the diffraction of a Gaussian profile is also Gaussian, the amplitude decreases with the distance from the source, but maintains a Gaussian shape. At a large distance $D \gg 0$ from the source, in the fundamental propagation mode, the wavefronts are almost spherical and the amplitude distribution on the mirror, $u(x_1, z_1)$, can be written as (Raimondi~et al.~\cite{RaimondiNIMA2013})
\begin{equation}
	u(x_1, z_1) =\sqrt{\frac{\Delta R_1}{\omega}\sqrt{\frac{2}{\pi}}}\exp\left[-\frac{(x_1-R_{\mathrm c})^2}{\omega^2}\right],
	\label{eq:FELprop}
\end{equation}
where $R_{\mathrm c} = R_0 + \Delta R_1/2$, assuming the beam to point toward the center of the mirror. The beam width rms, $\omega$, varies with $z_1$ according to the relation
\begin{equation}
	\omega(z_1) = \sqrt{\omega_0^2+\frac{\lambda^2 (D-z_1)^2}{\pi^2 \omega_0^2}},
	\label{eq:FELwaist}
\end{equation}
and $\omega_0$ is the beam width rms near the light source (beam ``waist''). The multiplicative constant in Eq.~(\ref{eq:FELprop}) is chosen to normalize the average beam intensity: 
\begin{equation}
	\frac{1}{\Delta R_1}\int_{-\infty}^{+\infty}\! |u(x_1, z_1)|^2 \,\mbox{d}x_1 = 1.
	\label{eq:FELnorm}
\end{equation}
Bendable mirrors can be used to turn the initial distribution of the beam (Eq.~(\ref{eq:FELprop})) into a desired one (Svetina~et al.~\cite{Svetina2012}), endowing the mirror with a properly designed profile (Spiga~et al.~\cite{Spiga2013}). Extension of the PSF equation, Eq.~(\ref{eq:PSF}), to the case of an anisotropic, coherent source is straightforward:
\begin{equation}
	\mbox{PSF}(x)=\frac{\Delta R_1}{L_1^2\lambda R_0}\left|\int_f^{f+L_1}\!\!\!\!\!\!u(x_1, z_1)\!\sqrt{\frac{x_1}{\bar{d}_{2,0}}}\,e^{-\frac{2\pi \mathrm{i}}{\lambda} \left[\bar{d}_{2,0}-z_1+\frac{x_1^2}{2(S-z_1)}\right]} \,\mbox{d}z_1\right|^2
	\label{eq:PSF_anis}
\end{equation}
which we explicitly solve in the next section for a perfect ellipsoidal mirror.

\subsection{Applications to the far-field configuration}\label{farfield}
In this section we simplify Eq.~(\ref{eq:PSF}) in the frequent case of an observation plane at a very large distance from the mirror, an approximation well-known as far-field (Fraunhofer) diffraction, retrieving known expressions based on the Fourier transform. We anticipate that this approximation cannot be applied to optical systems like the Wolter-I, in which two reflections occur in sequence at a short distance (Sect.~\ref{wolter}).

In the astronomical case, $S \rightarrow +\infty$, which means that the third term of the exponent in Eq.~(\ref{eq:PSF}) is negligible and $\alpha_1 = \alpha_0$. Hence, $\sin\alpha_0 = \Delta R_1/L_1$. Moreover, if the PSF is evaluated at the focal plane and $f\gg L_1$, then the square root in the integral varies much slower than the exponential and can be approximated by a constant, $\sqrt{R_0/f}$. Equation~(\ref{eq:PSF}) then reduces to
\begin{equation}
	\mbox{PSF}(x)=\frac{\Delta R_1}{L_1^2 \lambda f}\left|\int_f^{f+L_1}\!\!\! e^{-\frac{2\pi \mathrm{i}}{\lambda} \left[\sqrt{(x_1-x)^2+ z_1^2}-z_1\right]} \,\mbox{d}z_1\right|^2.
	\label{eq:PSF_far}
\end{equation}
We derived this simplified expression in a previous work (Raimondi \& Spiga~\cite{RaiSpi2010}). 

As a further step, we decompose the mirror profile as in Eq.~(\ref{eq:profile_comp}), where $x_{{\mathrm n}1}$ is a parabolic profile with the focus in the origin of the reference frame, and we denote the total profile error with $x_{{\mathrm e}1}$, in general below a micron of amplitude. Here $x_{{\mathrm n}1}$ plays the role of a ``lens'' that focuses at $z = 0$ the beam diffracted by $x_{{\mathrm e}1}$: in this way, the angular distribution of the PSF is solely determined by the entrance pupil size and by $x_{{\mathrm e}1}$ (alternatively, the beam can be initially converging and be diffracted by the error profile, Saha et al.~\cite{Saha2010}). We thereby write the expression under root in the exponent of Eq.~(\ref{eq:PSF_far}) as
\begin{equation}
	(x_1-x)^2+z_1^2 \simeq x^2+x_{{\mathrm n}1}^2+2x_{{\mathrm n}1}(x_{{\mathrm e}1}-x)+z_1^2,
	\label{eq:r_err}
\end{equation}
where we have neglected the term $x_{{\mathrm e}1}x$ because the focal spot and $x_{{\mathrm e}1}$ are usually much smaller than the mirror size. The Fraunhofer approximation consists of also neglecting the $x^2$ term, and developing the root at the first order. The exponent in Eq.~(\ref{eq:PSF_far}) becomes 
\begin{equation}
	\sqrt{(x_1-x)^2+z_1^2} -z_1\simeq \left[\sqrt{x_{{\mathrm n}1}^2+z_1^2}-z_1\right]-\frac{x_{{\mathrm n}1}(x-x_{{\mathrm e}1})}{\sqrt{x_{{\mathrm n}1}^2+z_1^2}}.
	\label{eq:PSF_far1}
\end{equation}
If one substitutes the equation of a parabola with the focus in the origin of the reference frame, $z_1 = ax^2_{{\mathrm n}1}-1/4a$, where $a$ is a positive constant, the term in [\,] brackets on right-hand side of Eq.~(\ref{eq:PSF_far1}) reduces to $1/2a$, a constant phase factor that can be ignored. Using this result, Eq.~(\ref{eq:PSF_far}) reduces to
\begin{equation}
	\mbox{PSF}(x)\simeq\frac{\Delta R_1}{L_1^2 \lambda f}\left|\int_f^{f+L_1}\!\!\! e^{-\frac{2\pi \mathrm{i}}{\lambda} \frac{x_{{\mathrm n}1}x}{z_1}}e^{-\frac{2\pi \mathrm{i}}{\lambda}x_{{\mathrm e}1}\, 2\sin\alpha}\,\mbox{d}z_1\right|^2,
	\label{eq:PSF_far2}
\end{equation}
where we approximated $\sqrt{x_{{\mathrm n}1}^2+z_1^2} \approx z_1$, always in far-field condition, and defined $2\sin\alpha \approx x_{{\mathrm n}1}/z_1$. Still, owing to the high value of $f$, $\alpha \simeq \alpha_0$, and using Eq.~(\ref{eq:rad_ampl}) we can also write $\mbox{d}z_1 \approx (L_1/\Delta R_1)\,\mbox{d}x_{{\mathrm n}1}$. 

We finally express the PSF as a function of the angular deviation defined in Sect.~\ref{isopoint}, $\theta = x/z_1 \approx x/f$, and find a well-known result:
\begin{equation}
	\mbox{PSF}(\theta)=\frac{1}{\Delta R_1 \lambda}\left|\int_{0}^{\infty}\! e^{-\frac{2\pi \mathrm{i}}{\lambda} \,x_{{\mathrm n}1} \theta}\, \mbox{CPF}(x_{{\mathrm n}1}) \,\mbox{d}x_{{\mathrm n}1}\right|^2,
	\label{eq:PSF_far3}
\end{equation}
where CPF($x_{{\mathrm n}1}$) denotes the complex pupil function:
\begin{equation}
	\mbox{CPF}(x_{{\mathrm n}1}) = \exp\left(-\frac{2\pi \mathrm{i}}{\lambda}\, 2 x_{{\mathrm e}1} \sin\alpha_0\right) ,
	\label{eq:CPF}
\end{equation}
which is zero outside the interval $[R_0, R_0+\Delta R_1]$. Equation~(\ref{eq:PSF_far3}) is the well-known expression of the far-field PSF, and the expression in the squared module -- the Fourier transform of the CPF -- is known as optical transfer function (OTF: Harvey~et al.~\cite{Harvey1988}).

A perfect mirror is represented by $x_{{\mathrm e}1} = 0$ everywhere: Eq.~(\ref{eq:PSF_far3}) then becomes
\begin{equation}
	\mbox{PSF}(\theta)=\frac{1}{\Delta R_1 \lambda}\left|\int_{R_0}^{R_0+\Delta R_1}\!\!\! e^{-\frac{2\pi \mathrm{i}}{\lambda} \,x_{{\mathrm n}1} \theta} \,\mbox{d}x_{{\mathrm n}1}\right|^2 = \frac{\beta}{\pi}\frac{\sin^2(\beta\theta )}{(\beta\theta)^2}
	\label{eq:PSF_perf}
\end{equation}
with $\beta = \pi \,\Delta R_1/\lambda$. Equation~(\ref{eq:PSF_perf}) is the expected diffraction pattern of a linear aperture of width $\Delta R_1$ (Fig.~\ref{fig:diffpat}). Moreover, it is correctly normalized to unity, as we anticipated in Sect.~\ref{isopoint}.

In real mirrors, $x_{{\mathrm e}1} \ne 0$. For example, if $x_{{\mathrm e}1}$ is a sinusoid (as in Sect.~\ref{singrat}), then Eq.~(\ref{eq:PSF_far3}) turns into
\begin{equation}
	\mbox{PSF}(\theta)=\frac{\Delta R_1}{L_1^2 \lambda}\left|\int_{f}^{f+L_1}\! e^{-\frac{2\pi \mathrm{i}}{\lambda} \,\sin\alpha_0 \left[z_1 \theta+2A\sin\left(\frac{2\pi z_1}{T}\right)\right]}\,\mbox{d}z_1\right|^2.
	\label{eq:PSF_far_sin}
\end{equation}
Reflectance maxima are located by Eq.~(\ref{eq:grat1}), which in shallow-angle approximation reads $T\theta_k \sin\alpha_0 \simeq k\lambda$ with $k$ integer, and the PSF at peaks becomes
\begin{equation}
	\mbox{PSF}(\theta_k) \approx \frac{\Delta R_1}{\lambda}\left|\frac{1}{2\pi}\int_{0}^{2\pi}\! e^{-\mathrm{i}\left(kt+\frac{4\pi A\sin\alpha_0}{\lambda}\sin t\right)}\,\mbox{d}t\right|^2,
	\label{eq:PSF_far_sin2}
\end{equation}
where we have set $t= 2\pi z_1/T$. In Eq.~(\ref{eq:PSF_far_sin2}), the expression in the square module is $J_k$, the $k^{th}$ Bessel function of the first kind, and we obtain the result anticipated in Eq.~(\ref{eq:grat2}):
\begin{equation}
	\mbox{PSF}(\theta_k) \approx \frac{\Delta R_1}{\lambda} \,J_k^2\left(\frac{4\pi A\sin\alpha_0}{\lambda}\right).
	\label{eq:PSF_far_sin3}
\end{equation}
Since every peak has a typical diffraction width of $\lambda/\Delta R_1$ and $\sum_{-\infty}^{+\infty} J_k^2(x) =1$ for any value of $x$, the PSF is correctly normalized. Equation~(\ref{eq:PSF_far_sin3}) is the shallow-angle approximation of the well-known diffraction pattern of a sinusoidal grating (see e.g., Stover~\cite{Stover95}).

More generally, decomposing $x_{{\mathrm e}1} = \sum_m x_m$ into different contributions (e.g., as measured with instruments sensitive to different windows of spatial frequencies, like in Eq.~(\ref{eq:profile_comp})), allows us to separate Eq.~(\ref{eq:CPF}) into the respective CPF factors:
\begin{equation}
	\mbox{CPF}(x_1) = \chi([R_0, R_0+\Delta R_1])\cdot\prod\nolimits_m \exp\left(-\frac{2\pi \mathrm{i}}{\lambda}\, 2 \, x_m \sin\alpha_0 \right)
	\label{eq:CPF_fact}
\end{equation}
where $\chi$ is the characteristic function of the interval $[R_0, R_0+\Delta R_1]$ and the $m^{th}$ profile error term is assumed to be infinitely extended. Since the contributions to the CPF are multiplicative, the respective transforms are to be convolved to return the total OTF. 

In far-field approximation, the OTF is thereby the convolution of the OTFs related to different components of the profile error, including the aperture diffraction term represented by the $\chi$ function. However, the same convolution is neither possible in near-field diffraction nor applicable to the squared module of the transform, that is, the PSF: it is therefore incorrect in general to convolve the PSFs of the different contributions to the profile error. For example, taking $x_{{\mathrm e}1} = x_{{\mathrm a}}+ x_{{\mathrm b}} $ with $x_{{\mathrm a}} = A\sin(2\pi z_1/T)$ and $ x_{{\mathrm b}} = -x_{{\mathrm a}}$, we have $x_{{\mathrm e}1} = 0$, the CPF reduces to the $\chi$ function and we correctly obtain Eq.~(\ref{eq:PSF_perf}). But, computing the PSF expected from $x_{{\mathrm a}}$ and $x_{{\mathrm b}}$ separately yields Eq.~(\ref{eq:PSF_far_sin3}) in both cases. The convolution of the two PSFs then returns a multiple peak structure very different from Eq.~(\ref{eq:PSF_perf}).

As a last example, we compute the PSF of a perfectly elliptical mirror ($x_{{\mathrm e}1} = 0$) illuminated by a distant FEL source in fundamental mode (Eq.~(\ref{eq:FELprop})) in far-field condition. By approximating the square root in the integrand with $\sqrt{R_0/f}$ as we did in Eq.~(\ref{eq:PSF_far}), Eq.~(\ref{eq:PSF_anis}) becomes 
\begin{equation}
	\mbox{PSF}(x)=\frac{\sqrt{2/\pi}}{\omega \lambda f}\left|\int_{R_0}^{R_0+\Delta R_1}\!\!\!\!\!e^{-\frac{(x_1-R_{\mathrm c})^2}{\omega^2}} e^{-\frac{2\pi \mathrm{i}}{\lambda} \left[\bar{d}_{2,0}-z_1+\frac{x_1^2}{2(S-z_1)}\right]} \,\mbox{d}x_1\right|^2:
	\label{eq:PSF_anis_far}
\end{equation}
developing the exponent in a similar way to Eq.~(\ref{eq:PSF_far1}), but this time substituting the equation of the ellipse, we can rewrite the previous equation as 
\begin{equation}
	\mbox{PSF}(x)=\frac{\sqrt{2/\pi}}{\omega \lambda f}\left|\int_{R_0}^{R_0+\Delta R_1}\!\!e^{-\left[\frac{(x_1-R_{\mathrm c})^2}{\omega^2}+2\frac{\pi\mathrm{i} x}{\lambda f} x_1\right]} \,\mbox{d}x_1\right|^2
	\label{eq:PSF_anis_far1}
\end{equation}
where the distance of the source to the mirror, $D$, is assumed to be large enough to take $\omega \simeq \lambda D/\pi \omega_0$ (Eq.~(\ref{eq:FELwaist})) approximately independent of $z$. Changing the integration variable to $t = (x_1-R_{\mathrm c})/\omega$, discarding unessential phase factors, and completing the square in the exponent, we obtain after some handling
\begin{equation}
	\mbox{PSF}(x)=\sqrt{\frac{2}{\pi}} \frac{D}{f\omega_0}e^{-2\left(\frac{Dx}{f\omega_0}\right)^2}\left|\frac{1}{\sqrt{\pi}}\int_{-\Delta R_1/2\omega}^{+\Delta R_1/2\omega}\!\!e^{-\left(t+\frac{\mathrm{i} D x}{f \omega_0} \right)^2} \,\mbox{d}t\,\right|^2.
	\label{eq:PSF_anis_far2}
\end{equation}
In Eq.~(\ref{eq:PSF_anis_far2}), the first exponential factor is the image of the Gaussian source, de-magnified by a factor $f/D$, whilst the complex error function in the square module accounts for the modulation caused by the Gaussian beam tail cutoff by the mirror aperture. If $\Delta R_1/2\omega \rightarrow \infty$, then the modulation factor tends to 1 and the PSF becomes exactly a Gaussian, as expected.

\section{Extension to a double-reflection optical system}\label{wolter}
In this section we extend the previous formalism to an optical system with two consecutive reflections such as a Wolter-I, a widespread optical system in X-ray astronomy, composed of two coaxial and confocal reflective surfaces: a paraboloid and a hyperboloid (Van Speybroeck \& Chase~\cite{VanSpeyChase}). However, different kinds of double-reflection systems are also adopted in X-ray astronomy, such as polynomial profiles (Conconi~et al.~\cite{Conconi2010}). For generality, we denote the two segments of the double-reflection system as primary and secondary mirror.
\begin{figure}[!tpb]
\centering
    \resizebox{\hsize}{!}{\includegraphics{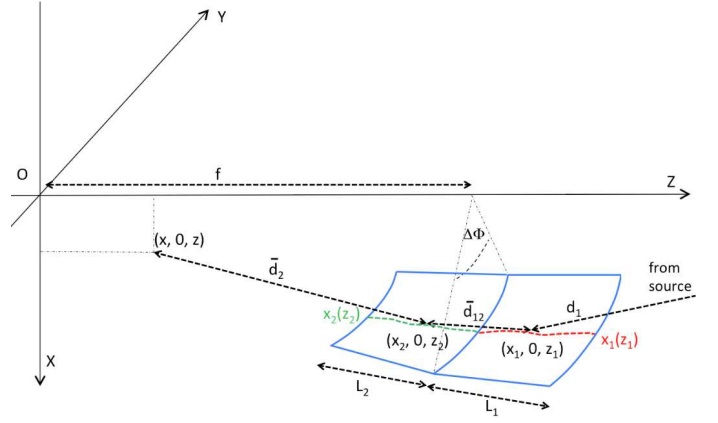}}
    \caption{Geometry of a double-reflection system such as a Wolter-I. The electric field on a meridional profile of the hyperbola is computed by Fresnel diffraction on the parabola. The PSF on the focal plane is subsequently computed.}
    \label{fig:Wolter}
\end{figure}

We thereby extend the optical setup as shown in Fig.~\ref{fig:Wolter}: we here also at first neglect azimuthal errors and assume that the profile is described by the radial coordinate as a function of $z$, which in the $xz$ plane is denoted with $x_1(z_1)$ of length $L_1$ for the primary, and $x_2(z_2)$ of length $L_2$ for the secondary. In the frequent case that $L_1 = L_2$, we denote their common value with $L$. For simplicity, the two mirror segments are assumed to intersect at $z =f$, even if an extension to an optical system with the two segments separated by a gap is straightforward. The primary mirror collects the radiation from an isotropic, point-like X-ray source at $z = S$ and diffracts it onto the secondary, which eventually diffracts the wave to a focus. The nominal focal plane is still assumed to be at $z =0$. The angle formed by the two surfaces at the intersection plane is $2\alpha_0$, and $R_0$ is the corresponding azimuthal curvature radius. Finally, always denoting with $\delta = R_0/D$ the beam divergency (negative for a converging wave), the incidence angles is $\alpha_1 = \alpha_0+\delta$ on the primary segment and $\alpha_2 = \alpha_0-\delta$ on the secondary (Spiga~et al.~\cite{Spiga2009}). The corresponding radial amplitudes are $\Delta R_1 = L_1\alpha_1$ and $\Delta R_2 = L_2\alpha_2$. Clearly, for an on-axis astronomical source we have $\alpha_1 = \alpha_2$: if additionally we have $L_1 = L_2$, then we also have $\Delta R_1 = \Delta R_2$.

\begin{figure}[!tpb]
	\resizebox{\hsize}{!}{\includegraphics{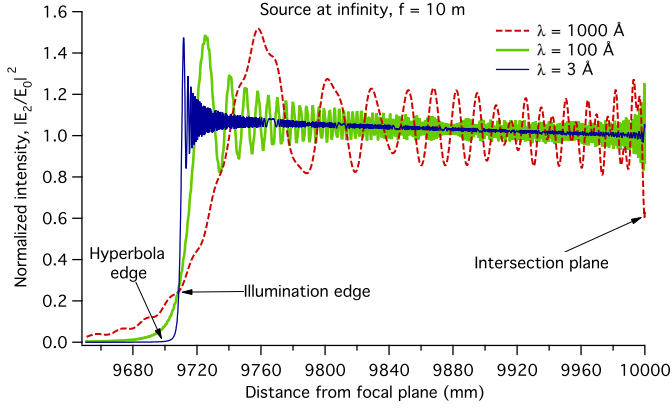}}
    \caption{Field intensity along the hyperbolic profile in a perfect Wolter-I mirror at three different values of $\lambda$, as computed with Eq.~(\ref{eq:field2}). We assumed that $L_1 = L_2$ = 300~mm, $R_0$ = 150~mm, $f$~= 10~m, and $D \rightarrow \infty$.}
    \label{fig:field_hyp1}
\end{figure}

The electric field diffracted by the primary mirror (Fig.~\ref{fig:Wolter}) can be computed on the profile of the secondary mirror using Eq.~(\ref{eq:field}) for all the points of the secondary mirror: 
\begin{equation}
	E_2(x_2,z_2)=\frac{E_0\, \Delta R_1}{L_1\sqrt{\lambda x_2}}\int_f^{f+L_1}\!\!\!\!\sqrt{\frac{x_1}{\bar{d}_{12}}}\,e^{-\frac{2\pi \mathrm{i}}{\lambda} \left[\bar{d}_{12}-z_1+\frac{x_1^2}{2(S-z_1)}\right]} \,\mbox{d}z_1,
	\label{eq:field2}
\end{equation}
where $\bar{d}_{12}$ is the distance in the $xz$ plane from a generic point of the primary mirror to a generic point on the secondary mirror:
\begin{equation}
	\bar{d}_{12} = \sqrt{(x_2-x_1)^2+ (z_2-z_1)^2}.
	\label{eq:d12_av_doub}
\end{equation} 
An example of applying of Eq.~(\ref{eq:field2}) to a Wolter-I perfect profile is shown in Fig.~\ref{fig:field_hyp1} at three different values of $\lambda$, in the most frequent configuration for astronomical mirrors: source at infinity, on-axis, and $L_1 = L_2$, and then also $\Delta R_1 = \Delta R_2$. The normalized field intensity, $|E_2/E_0|^2$, is computed over the hyperbola length and 50 mm beyond to show the extension of the diffracted field.

The electric field intensity on the hyperbolic profile exhibits the characteristics of the Fresnel diffraction pattern from a straight edge (Fig.~\ref{fig:field_hyp1}). Even if the two segments have the same length and incidence angle, the region geometrically illuminated by the parabolic segment is slightly shorter than the hyperbola length because of the parabolic mirror axial curvature. At the illumination edge, the intensity is always one quarter of the incident intensity, then decreases gradually in the geometric shaded region. In the illuminated region, the intensity is modulated by diffraction fringes of increasing frequency as $\lambda$ decreases. Exactly like the example in Sect.~\ref{singrat}, the results tend to the geometrical optics findings in the limit of low $\lambda$ values. Finally, the increasing intensity from the intersection plane toward the illumination edge denotes the progressive power concentration, as expected from a focusing mirror.

\begin{figure}[!tpb]
	\resizebox{\hsize}{!}{\includegraphics{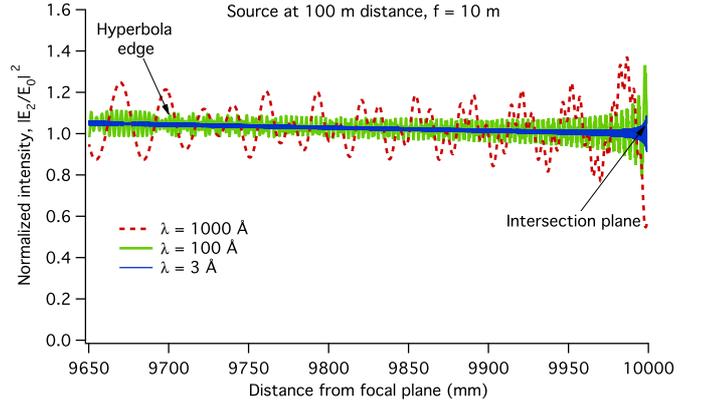}}
    \caption{Field intensity along the hyperbolic profile in a perfect Wolter-I mirror as computed with Eq.~(\ref{eq:field2}). Same mirror parameter values as in Fig.~\ref{fig:field_hyp1}, but this time $D=$ +100~m.}
    \label{fig:field_hyp2}
\end{figure}

However, the situation changes if the source is at finite distance: if $\delta > 0$ we already expect from geometrical optics that a fraction of rays reflected by the primary mirror miss the second reflection, and the effective radial aperture is reduced from $\Delta R_1$ to $\Delta R_2$. Vice versa, if $\delta < 0$, all rays undergo the second reflection, but the radial aperture of the primary mirror is reduced to $\Delta R_1$. Hence, the effective radial aperture for a double reflection for a source on-axis can be shortly written as
\begin{equation}
	\Delta R_{\mathrm{m}} = \min(\Delta R_1, \Delta R_2),
	\label{eq:rmin}
\end{equation} 
provided that it is non-negative (Spiga~et al.~\cite{Spiga2009}). Applying Eq.~(\ref{eq:field2}) to a source at finite distance returns a similar picture. For example, we report in Fig.~\ref{fig:field_hyp2} the computation of the normalized intensity for a source with $D$~= 100~m: even if the diffraction pattern is still visible, the illumination edge is not, at least within the hyperbola length: this means that part of the wavefront after the first reflection is diffracted beyond the edge of the secondary. Hence, a relevant amount of the collected power is lost in single reflection, in accord with geometrical optics expectations. The modulation visible in Figs.~\ref{fig:field_hyp1} and~\ref{fig:field_hyp2} is, however, a typical physical optics effect.

The subsequent diffraction by the secondary segment, at any position in the $xz$ plane (in-, intra-, or extra-focus), is simply obtained from applying Eq.~(\ref{eq:field}) weighting its integrand on the complex $E_2$ function obtained from Eq.~(\ref{eq:field2}):
\begin{equation}
	E(x, z)=\frac{\Delta R_2}{L_2\sqrt{\lambda x}}\int^f_{f-L_2}\!\!\!\!E_2(x_2, z_2)\sqrt{\frac{x_2}{\bar{d}_{2}}}\,e^{-\frac{2\pi \mathrm{i}}{\lambda} \bar{d}_2} \,\mbox{d}z_2.
	\label{eq:field3}
\end{equation}
In the last equation the complex expression of $E_2$ already includes all the relevant information on the phase; hence, the terms in the exponent that include subscript 1 have been removed. Only the distance $\bar{d}_{2}$ remains:
\begin{equation}
	\bar{d}_{2} = \sqrt{(x_2-x)^2+ (z_2-z)^2}.
	\label{eq:d2_av_doub}
\end{equation}  
Finally, the computation of the PSF in the nominal focal plane is made taking the squared module of Eq.~(\ref{eq:field3}) at $z =0$, and normalizing to the intensity collected within the radial aperture effective for double reflection, $\Delta R_{\mathrm{m}}$ (Eq.~(\ref{eq:rmin})):
\begin{equation}
	\mbox{PSF}_2(x) =\frac{(\Delta R_2)^2}{E_0^2 \Delta R_{\mathrm{m}}L_2^2 \,\lambda R_0}\left| \int_{f-L_2}^{f} \!\!\!\! E_2(x_2, z_2)\sqrt{\frac{x_2}{\bar{d}_{2,0}}}\, e^{-\frac{2\pi\mathrm{i}}{\lambda} \bar{d}_{2,0}} \, \mbox{d}z_2\right|^2
	\label{eq:PSF_W1}
\end{equation}
where $\bar{d}_{2,0}$ is Eq.~(\ref{eq:d2_av_doub}) evaluated at $z =0$. The last expression is independent of the incident radiation intensity, and normalized to 1 when integrated over $x$. 

Exactly as in Sect.~\ref{isopoint}, we have to set an appropriate sampling of the primary mirror profile, of the secondary mirror profile, and of the focal line. For the secondary mirror sampling $\Delta z_2$, just replacing $\alpha_1 \rightarrow \alpha_2$ in Eq.~(\ref{eq:minsamp_z}) is necessary (Eq.~(\ref{eq:minsamp_z2})). For the focal line sampling $\Delta x$, Eq.~(\ref{eq:minsamp_x}) is used changing $\alpha_1 \rightarrow \alpha_2$, $L_1 \rightarrow L_2$, and we have Eq.~(\ref{eq:minsamp_x2}). Finally, replacing the angle subtended by the detector in Eq.~(\ref{eq:minsamp_z}) with the angle subtended by the secondary mirror, we obtain Eq.~(\ref{eq:minsamp_z1}): 
\begin{eqnarray}
	\Delta z_1 &= &\frac{\lambda }{8\pi\alpha_0 \sin\alpha_1}\left(1+\frac{L_1}{L_2}\right) \label{eq:minsamp_z1}\\
	\Delta z_2 &= &\frac{\lambda f}{4\pi \sin\alpha_2\,\rho} \label{eq:minsamp_z2}\\
	\Delta x &=& \frac{\lambda f}{2\pi \sin\alpha_2\,L_2}.\label{eq:minsamp_x2}
\end{eqnarray}
As for the single-reflection case, applying Eqs.~(\ref{eq:field2}) and~(\ref{eq:PSF_W1}) with the sampling values provided by Eqs.~(\ref{eq:minsamp_z1}) to~(\ref{eq:minsamp_x2}) enables computing the PSF for any value of $\lambda$ within the detector field; including higher measured frequencies is always possible and results in an enhanced scattering out of the detector field and in a lower PSF normalization. Misalignments, offsets, and tilts of the two mirror segments can be included in the two profiles $x_1(z_1)$, $x_2(z_2)$ to be accounted for in the calculation.
\begin{figure}[!tpb]
\centering
    \subfigure[]{\includegraphics[width=0.5\textwidth]{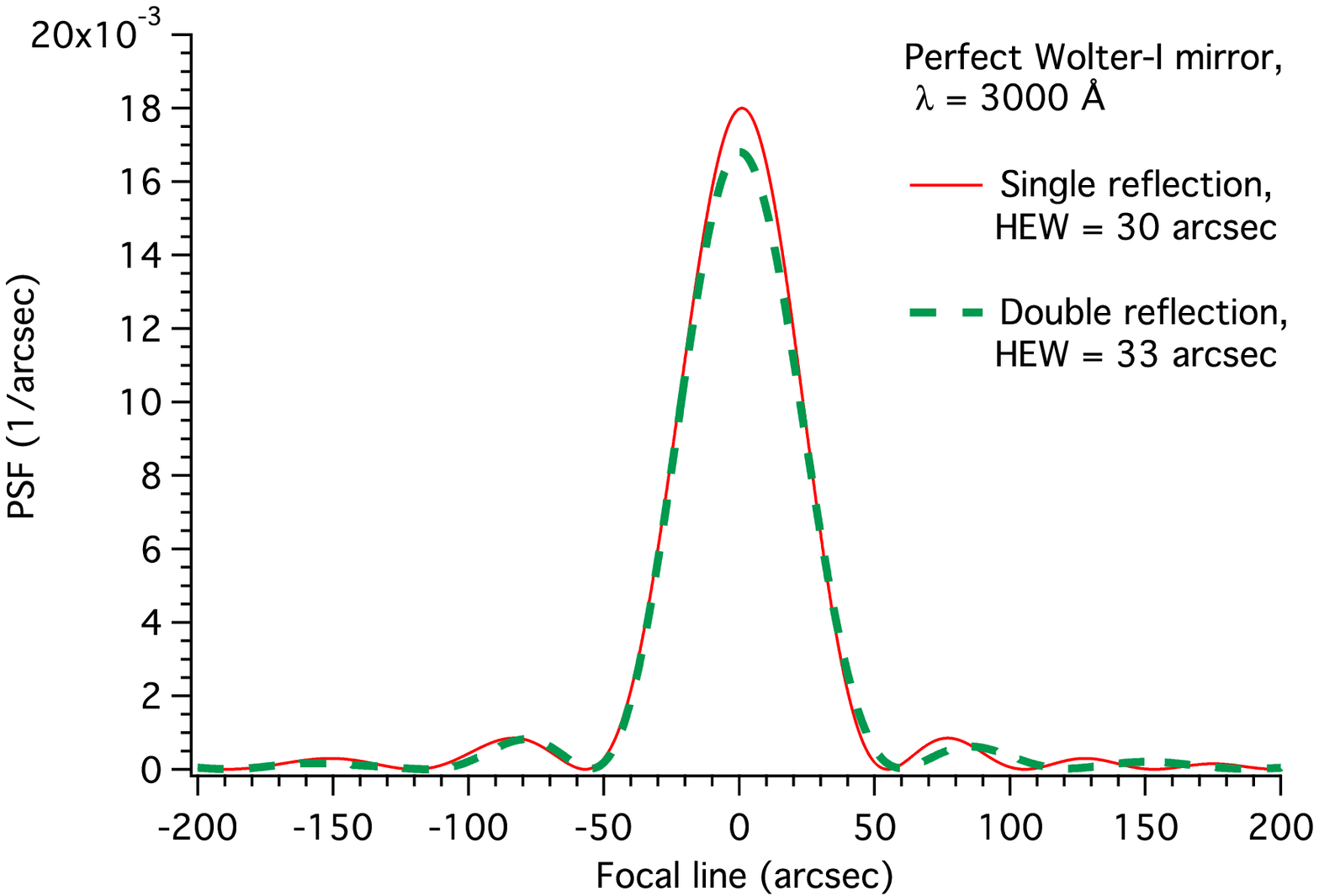}    \label{fig:perfect_mirror_a}}\\
    \subfigure[]{\includegraphics[width=0.5\textwidth]{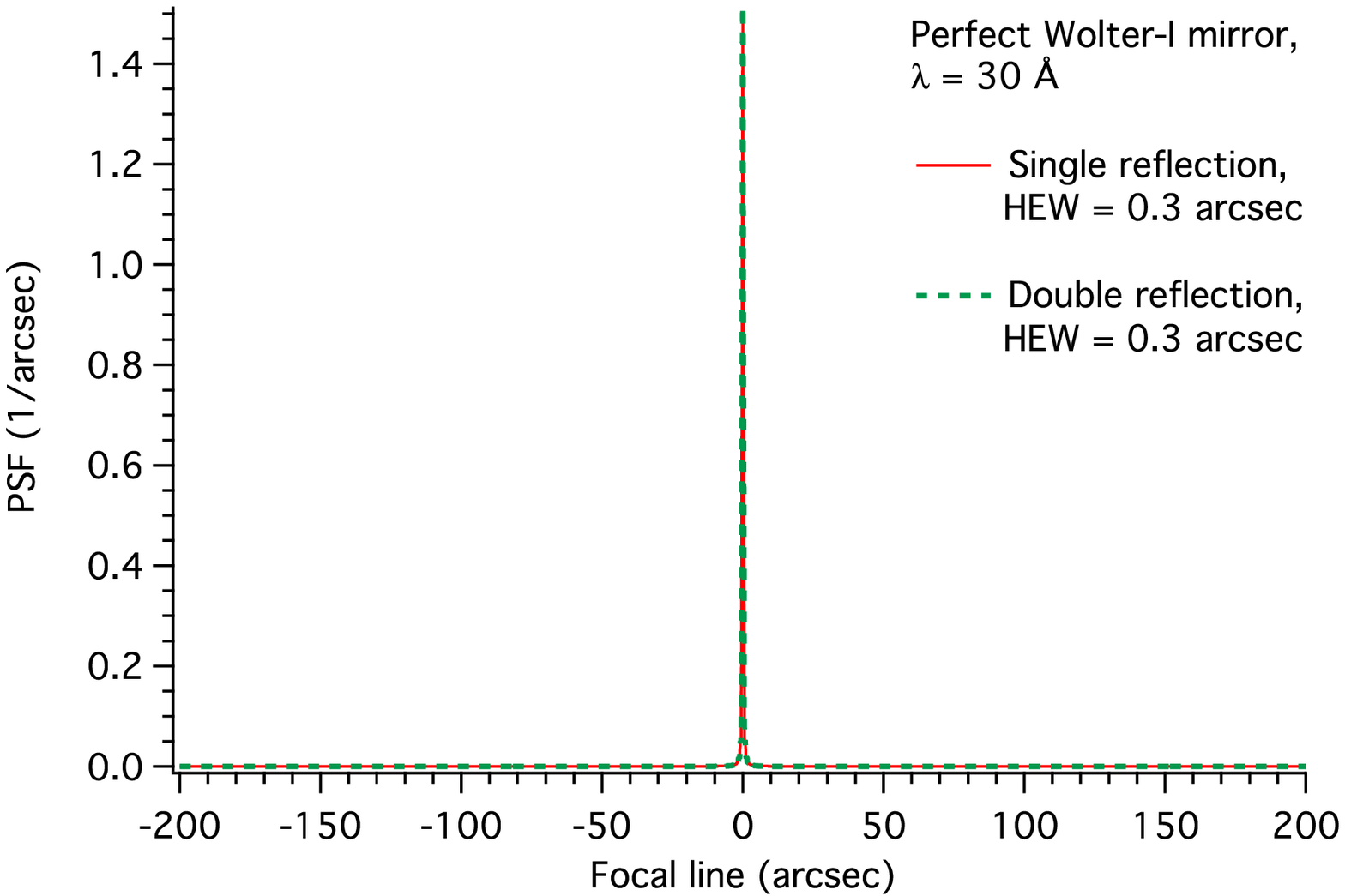}    \label{fig:perfect_mirror_b}}
    \caption{PSF of a perfect Wolter-I mirror with the same geometry as in Fig.~\ref{fig:field_hyp1} as computed with the WISE code. (a) in UV light at 3000~\AA, in single and double reflection: the aperture diffraction is apparent and enhanced in double reflection. (b) at 30~\AA, the aperture diffraction is minimized and the PSF becomes a Dirac delta, also in double reflection.}
\end{figure}

Equations~(\ref{eq:field2}) and~(\ref{eq:PSF_W1}) can be applied with the sole restrictions that the incidence angle is shallow and that the out-of-plane deflection effect of azimuthal errors are negligible. In particular, the approximations required in far-field condition after the first reflection (Sect.~\ref{farfield}) become redundant. On the other hand, the far-field condition cannot be applied when the two segments are continuous or separated by a few millimeters, as in the Wolter-I design, even if $\lambda$ is much smaller than the distances at play. The far-field approximation requires -- inter alia -- approximating the root in Eq.~(\ref{eq:field2}) with a constant, but this cannot be done because $d_{12}$ varies from $\sim L_1+L_2$ down to near zero. Such a rough approximation would make the diffraction pattern very different from the pattern correctly described in Figs.~\ref{fig:field_hyp1} and~\ref{fig:field_hyp2}. 

The far-field approximation is sometimes used to compute the double-reflection PSF using the CPF transform (Eq.~(\ref{eq:PSF_far3})), assuming as profile error the sum of the defects of the two segments: this implies the initial assumption that the wavefront at the mirror exit has a uniform intensity and a phase shift equal to the superposition of the phase shifts caused by the two profile errors. This in general incorrect, however, because Fig.~\ref{fig:field_hyp1} shows that the intensity on the secondary segment is nonuniform. Hence, using Eq.~(\ref{eq:PSF_far3}) for a Wolter-I system may lead to inaccurate results. We show this along with other examples in Sect.~\ref{longper}.

We finally mention that the extension to an extended or/and anisotropic source can be obtained in a completely analogous way to the one described in Sect.~\ref{incoh}.

\section{Examples of PSF computation for Wolter-I mirrors}\label{PSFcomp}
In the remainder of this paper we make use of Eqs.~(\ref{eq:field2}) and~(\ref{eq:PSF_W1}) to simulate the PSFs of Wolter-I mirrors characterized by profile errors and roughness. To this end, we have written a numerical code in IDL language named WISE to numerically solve the integrals, and in this section we show some results to demonstrate the versatility of the method in use. We show that all the computed PSFs behave as expected, accounting simultaneously for aperture diffraction, geometric errors, and roughness, without needing to adopt different treatments depending on the frequencies of the error profile. Some results for a Wolter-I mirror were anticipated in Raimondi \& Spiga~(\cite{RaiSpi2011}). Some applications of WISE to a Kirkpatrick-Baez optical system in use at FERMI at Elettra have been presented in Raimondi~et al.~(\cite{RaimondiNIMA2013}, \cite{RaimondiSPIE2013}). 

The test case we consider here is the Wolter-I mirror shell with parameters listed in the caption of Fig.~\ref{fig:field_hyp1}, adding different types of profile errors superimposed on either one or both segments of the nominal Wolter-I profile. 

\subsection{Perfect parabola and hyperbola}\label{perf}
The analytical expressions $x_1(z_{\mathrm 1})$ and $x_{\mathrm 2}(z_{\mathrm 2})$ of a Wolter-I nominal profile (Van Speybroeck \& Chase~\cite{VanSpeyChase}), when substituted into Eqs.~(\ref{eq:field2}) and~(\ref{eq:PSF_W1}) return a sinc-shaped PSF that becomes more peaked and narrower as $\lambda$ is decreased, as expected. The situation is completely analogous to the single reflection of a parabolic mirror (Eq.~(\ref{eq:PSF_perf})). In UV light, the broadening caused by the aperture diffraction is clearly seen in both single and double reflection and dominates the HEW value (Fig.~\ref{fig:perfect_mirror_a}). One might expect the HEW in Wolter-I configuration to be larger than the configuration resulting from the sole perfect parabolic segment because the wavefront was diffracted twice, and the result is in accord with the expectation. The interpretation is that the wavefront has become divergent after the first diffraction and becomes enlarged before impinging onto the secondary mirror, which in turn diffracts it by trimming its edges. In contrast, the result would have been indistinguishable from a single diffraction if computed from the product of the two segments' CPFs (Sect.~\ref{farfield}). Since optical tests on Wolter-I X-ray mirrors are performed in UV or visible light, the accurate subtraction of the diffraction aperture term should also account for the small difference introduced by the double reflection. 

In X-rays (0.4~keV, Fig.~\ref{fig:perfect_mirror_b}), the aperture diffraction is usually reduced to negligible levels and the PSF resembles a Dirac delta function for single- and double-reflection cases, as expected. The low but finite HEW value is determined by the spatial resolution of the focal line (10~$\mu$m). High-quality X-ray optics, however, can reach a PSF very close to the aperture diffraction limit.

\subsection{Sinusoidal grating on parabola, perfect hyperbola}\label{grating}
\begin{figure}[!btp]
\centering
    {\includegraphics[width=0.5\textwidth]{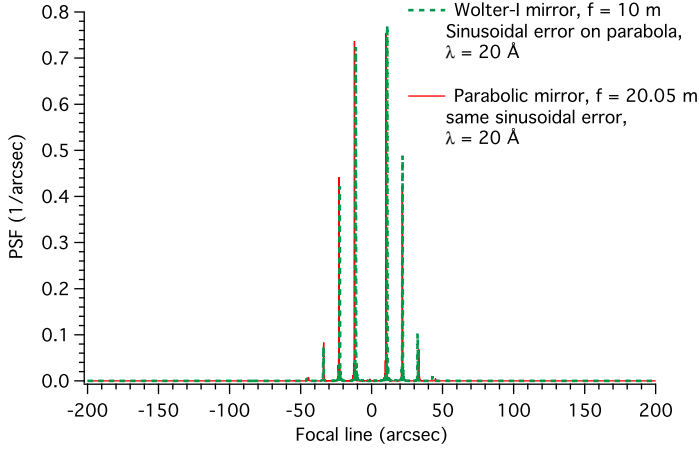}}
    \caption{PSF at $\lambda$ = 20~\AA~of a Wolter-I mirror with the same dimensions as in Fig.~\ref{fig:field_hyp1}, plus a sinusoidal perturbation (period 10~mm, amplitude 0.1~$\mu$m) on the sole parabola. The result is compared with the PSF of the sole parabolic mirror segment with the same defect, detected in the focal plane of the parabola at a distance of approx. 20~m. The PSF simulation for the Wolter-I mirror with the perturbed parabolic profile returns the same result as the single-reflection case.}
    \label{fig:grating_double}
\end{figure}

As a first example of an imperfect Wolter-I mirror (sized as in the caption of Fig.~\ref{fig:field_hyp1}), we have considered a sinusoidal perturbation with an amplitude of of 0.1~$\mu$m and a period of 10~mm, superposed on the sole parabolic profile (Fig.~\ref{fig:grating_double}). This case was already treated extensively -- at a different incidence angle -- for a single-reflection mirror in Sect.~\ref{singrat}. The computed PSF at $\lambda$ = 20~\AA~exhibits the characteristically peaked pattern of a sinusoidal grating here as well (Eq.~(\ref{eq:PSF_far_sin3})). 

Since the hyperbola profile is not perturbed and the aperture diffraction effects are negligible at this value of $\lambda$, we expect that the beam diffracted by the sinusoidal grating is simply reflected to the focal plane, preserving the intensity distribution. In fact, the simulated PSF is very well superposed on the PSF of the sole perturbed parabolic profile (Fig.~\ref{fig:grating_double}: the focal length of the parabola slightly exceeds twice the focal length of the corresponding Wolter-I mirror). This example puts Eqs.~(\ref{eq:field2}) and~(\ref{eq:PSF_W1}) to the test: if the calculation were inaccurate, the second diffraction would not have reproduced the positions and heights of the single-reflection peaks, which in turn are confirmed by a comparison with the findings of the grating theory (Sect.~\ref{singrat}). The same result can be obtained by imparting the sinusoidal error to the sole hyperbola.

\subsection{Long-period deformations of parabola and hyperbola}\label{longper}
We now consider a deformation on both mirror segments over a lateral scale equal to $L$, whose effects are expected to merge at sufficiently low $\lambda$ values (e.g., 20~\AA) with the findings of a ray-tracing routine. Instead of a sinusoidal perturbation we have adopted the following profile error, superposed on the two segments of the Wolter-I profile: 
\begin{equation}
	x_{\mathrm{meas},j}(z_j) = -\frac{L_jw_j}{4\pi}\log\cos\left[\frac{\gamma \,(z_j-z_{0,j})}{L_j}\right],
	\label{eq:shape}
\end{equation}
where $j =1,2$, $z_{0,j}$ is the central coordinate of the $j^{th}$ segment, and $\gamma$ must be slightly smaller than $\pi$ to avoid the profile divergence at the edges. This profile error is specifically designed (Spiga et al.~\cite{Spiga2013}) to return in single reflection a Lorentzian-shaped PSF of HEW $|w_1|$, if geometrical optics can be applied, 
\begin{equation}
	\mbox{PSF}(\theta) = \frac{2|w_1|}{\pi \left(w_1^2+4\theta^2\right)},
	\label{eq:king}
\end{equation}
where $\theta = x/f$. A Lorentzian function like this is a special case of the more general King function, a realistic model for the PSF of an X-ray telescope (e.g., SWIFT-XRT's: Moretti et al.~\cite{Moretti2004}). To simulate the deformation effect on a Wolter-I mirror, we assume $x_{\mathrm{meas}1}(z_1)$ and $x_{\mathrm{meas}2}(z_2)$ to comply with Eq.~(\ref{eq:shape}). We then compute the PSFs in focus by applying Eqs.~(\ref{eq:field2}) and~(\ref{eq:PSF_W1}). 
\begin{figure}[!tpb]
\centering
    {\includegraphics[width=0.40\textwidth]{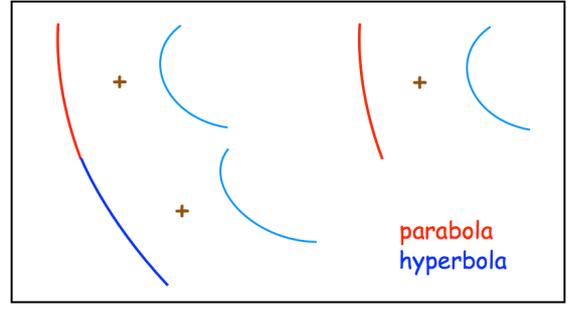}}
    {\includegraphics[width=0.5\textwidth]{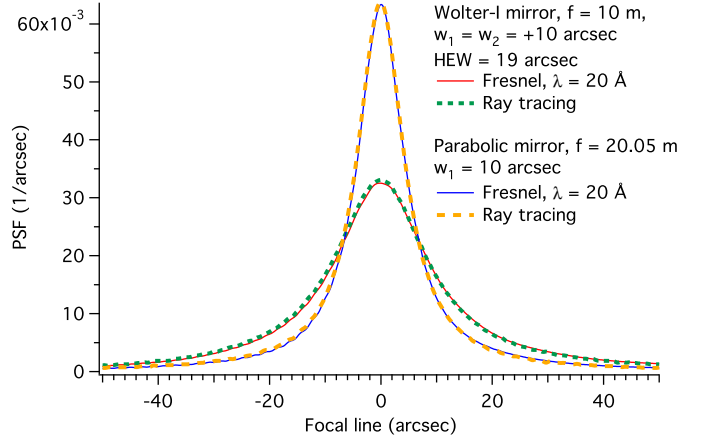}}
    \caption{Results of WISE for a Wolter-I mirror with the same geometrical properties as in Fig.~\ref{fig:field_hyp1} and a profile error according to Eq.~(\ref{eq:shape}) at $\lambda$~=20~\AA, assuming $w_1 = w_2 >0$. Also plotted for comparison is the PSF computed in single reflection (Eq.~(\ref{eq:PSF})), which correctly fits Eq.~(\ref{eq:king}). At this $\lambda$ and with this profile error, the results of physical optics (solid lines) accurately match the ray-tracing prediction (dashed lines).}
    \label{fig:double_curvature}
\end{figure}

\begin{figure}[!tpb]
\centering
    {\includegraphics[width=0.40\textwidth]{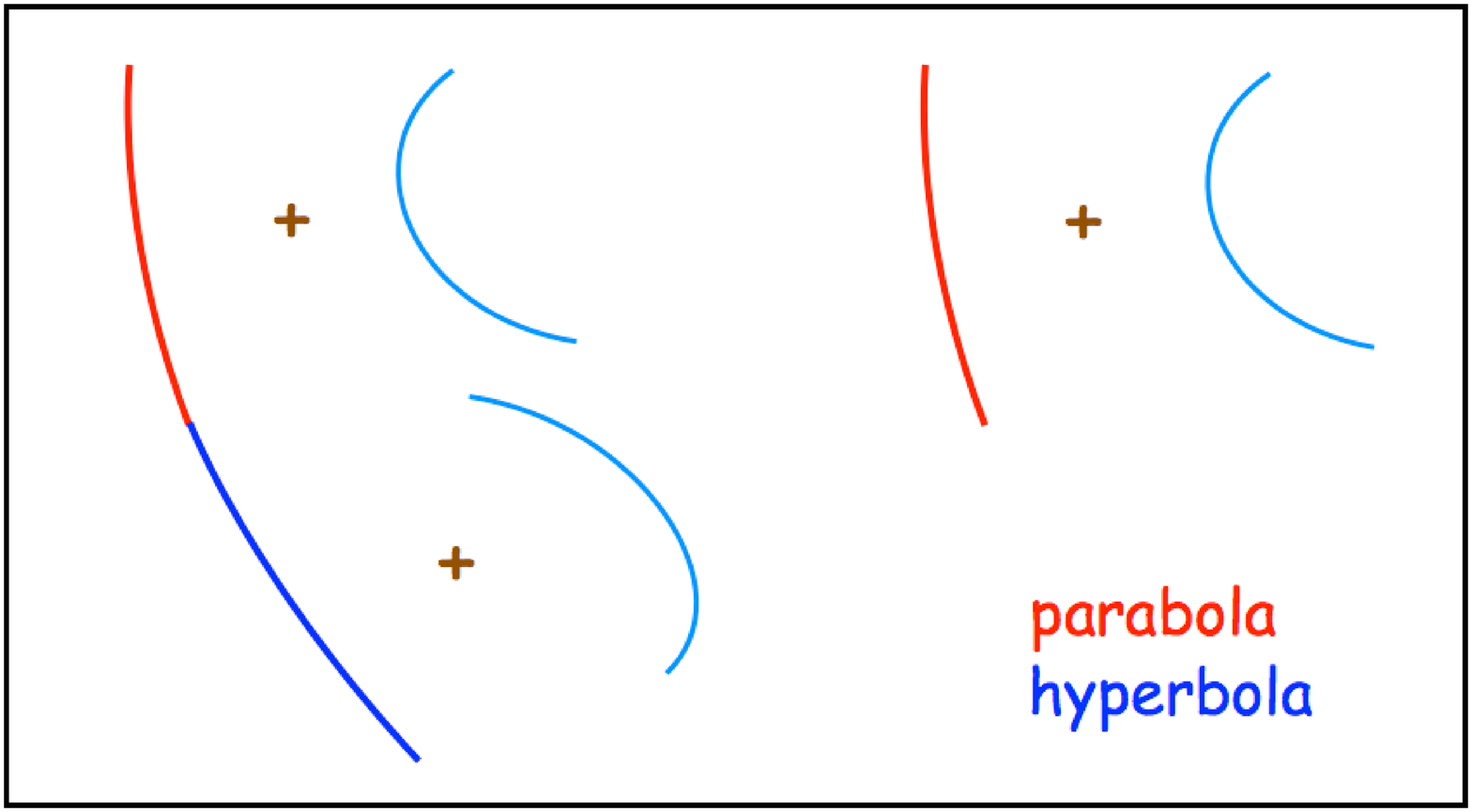}}
    {\includegraphics[width=0.5\textwidth]{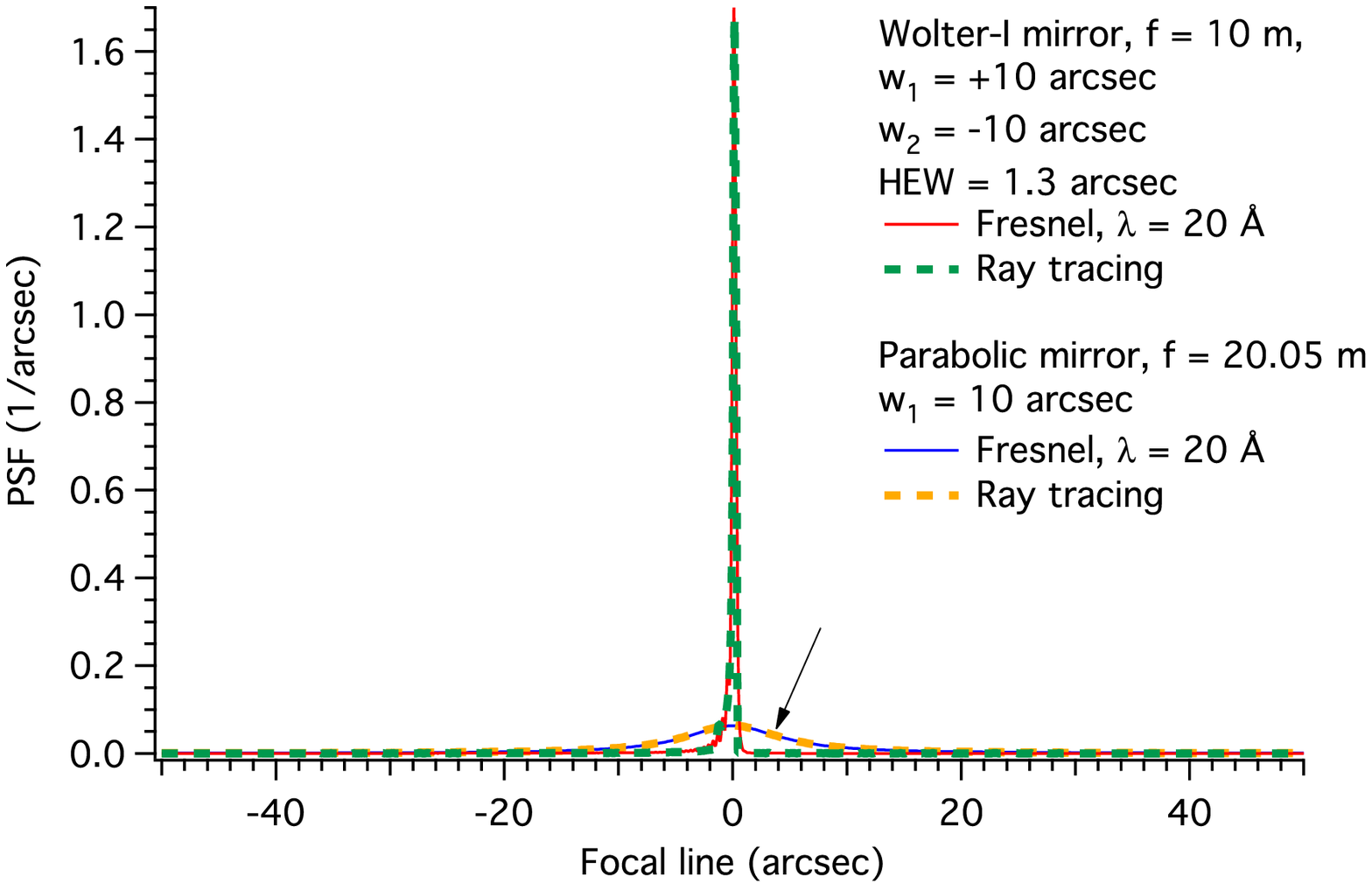}}
    \caption{WISE results for a Wolter-I mirror as in Fig.~\ref{fig:double_curvature} at $\lambda$~= 20~\AA~with the profile error described by Eq.~(\ref{eq:shape}) and $w_1 >0$, but this time $w_2 < 0$. Also plotted for comparison is the PSF simulated from the single reflection (arrow). The opposite curvatures of the profile errors largely balance each other, as expected. Here the predictions of the Fresnel diffraction (lines) also agree with the geometrical optics (dots).}
    \label{fig:opposite_curvature}
\end{figure}

If the two deformations are concave upward (i.e., $w_1$ and $w_2$ are both positive), the PSF spread is amplified with respect to the single-reflection case (Fig.~\ref{fig:double_curvature}). More exactly, the HEW of the deformed mirror equals 19~arcsec, roughly twice the HEW in single reflection. In contrast, if the error concavity is upward for the parabola and downward for the hyperbola (Fig.~\ref{fig:opposite_curvature}), the PSF computation correctly returns a quasi-delta function, with an HEW much smaller than the HEW of the single-reflection mirrors because the angular deviations approximately compensate for each other. As initially expected, this result also perfectly agrees with the results of a ray-tracing routine. 

\begin{figure}[!tpb]
\centering
    {\includegraphics[width=0.5\textwidth]{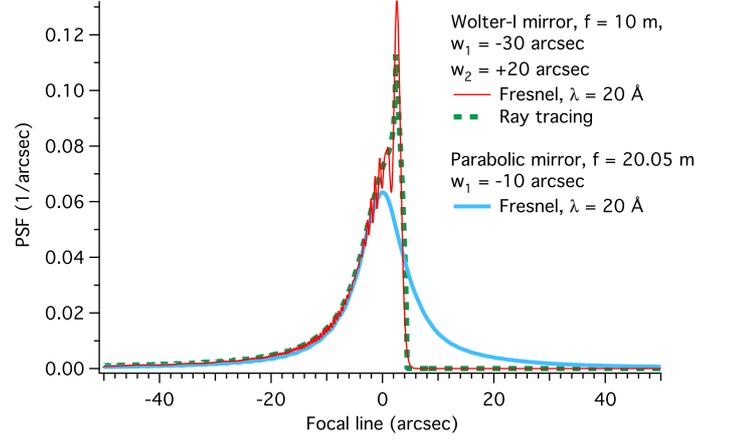}}
    \caption{WISE results for a Wolter-I mirror, $\lambda$~= 20~\AA,~with the dimensions listed in Fig.~\ref{fig:double_curvature} and the same kind of error, but now $w_1$= -30~arcsec, $w_2$= +20~arcsec. The PSF computed from Eqs.~(\ref{eq:field2}) and~(\ref{eq:PSF_W1}) now has a pronounced asymmetry, confirmed by the ray-tracing findings (dashes). The detector resolution is 50~$\mu$m. In contrast, the single-reflection PSF computed assuming a profile error according to Eq.~(\ref{eq:shape}) with $w_1$ = -10~arcsec would return a completely incorrect picture (blue line).}
    \label{fig:diff_curvature}
\end{figure}

It is worth noting that this example disproves the widespread misconception of a Wolter-I mirror PSF obtained as a convolution of the PSFs of the two mirror segments. However, it is also easy to show that adding the deformations of the two segments and applying the single-reflection formulae to the sum leads to inaccurate results. For example, Fig.~\ref{fig:diff_curvature} reports the double-reflection PSF simulated for concavities of different sign, but with $|w_1| > |w_2|$. Because the two profile errors have signs with opposite concavities but different amplitudes, the angular deviations do not exactly compensate for each other like in Fig.~\ref{fig:opposite_curvature}, and the PSF still has a finite width. However, the PSF predicted from applying Eqs.~(\ref{eq:field2}) and~(\ref{eq:PSF_W1}) now is clearly asymmetric, in turn confirmed by the ray-tracing (Fig.~\ref{fig:diff_curvature}). In contrast, if the PSF were computed using the single-reflection formalism -- for example, aiming at using the far-field formula (Eq.~(\ref{eq:PSF_far3})) applied to the sum of the two profile errors -- then a completely different, perfectly symmetric PSF would be obtained (blue solid line in Fig.~\ref{fig:diff_curvature}). 

The correct PSF is asymmetric, however. The reason is that the parabolic mirror error is concave downward -- unlike the example in Fig.~\ref{fig:opposite_curvature} -- and causes a divergence of the diffracted beam that impinges on the hyperbola broader than it initially was. Part of the wavefront is diffracted to the focal plane, but the remainder misses the hyperbola beyond the edge at $z = f-L_2$ and is lost. The resulting PSF must therefore exhibit a cutoff that is not observed in the single reflection PSF, but is correctly present in the two computations using the Fresnel diffraction and ray-tracing. Moreover, the two asymmetric PSFs have a normalization reduced to 87\%, consistent with a power loss in double reflection. However, they differ in one point: the former exhibits small oscillations near the cutoff that are not seen in the latter and therefore have to be diffraction fringes. The diffraction occurs exactly at the edge of the hyperbolic mirror, giving rise to fringes that, according to the theory of diffraction off a straight edge, are $\sqrt{\lambda/f} \simeq$ 3~arcsec wide near the cutoff. This number is in good accord with the observed fringes in the red line of Fig.~\ref{fig:diff_curvature}.

\begin{figure}[!tpb]
\centering
    {\includegraphics[width=0.5\textwidth]{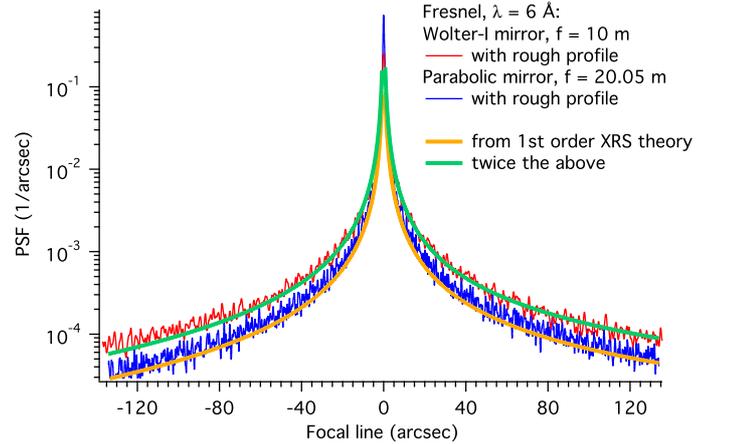}}
    \caption{Results of the WISE code for a Wolter-I mirror with the same characteristics as in Fig.~\ref{fig:field_hyp1}, at $\lambda = 6~\AA$, but rough profiles (semi-log scale). Roughness profiles generated from a power-law PSD (Eq.~(\ref{eq:powerlaw})) with $n$ = 1.5, $K_n$~= 150~nm$^3~\mu$m$^{-1.5}$ were superposed on the parabolic and hyperbolic segment. We also traced for $|\theta| >$ 1~arcsec the X-ray scattering diagram computed from the PSD via the first-order scattering theory. The agreement is excellent.}
    \label{fig:powlaw}
\end{figure}

\subsection{Parabola and hyperbola with roughness}\label{rough}
We now consider a Wolter-I mirror of the same dimensions as in the previous sections, with a surface roughness described by a power spectral density (PSD), to compare our formalism with the well-consolidated first-order scattering theory, therefore assuming that the roughness fulfills the smooth-surface limit condition (Eq.~(\ref{eq:roughlim})). In this way, the PSF should only depend on the chosen PSD and not on the exact rough profile realization.
\begin{figure*}[!tpb]
	\centering
     \subfigure[]{\includegraphics[width=0.45\textwidth]{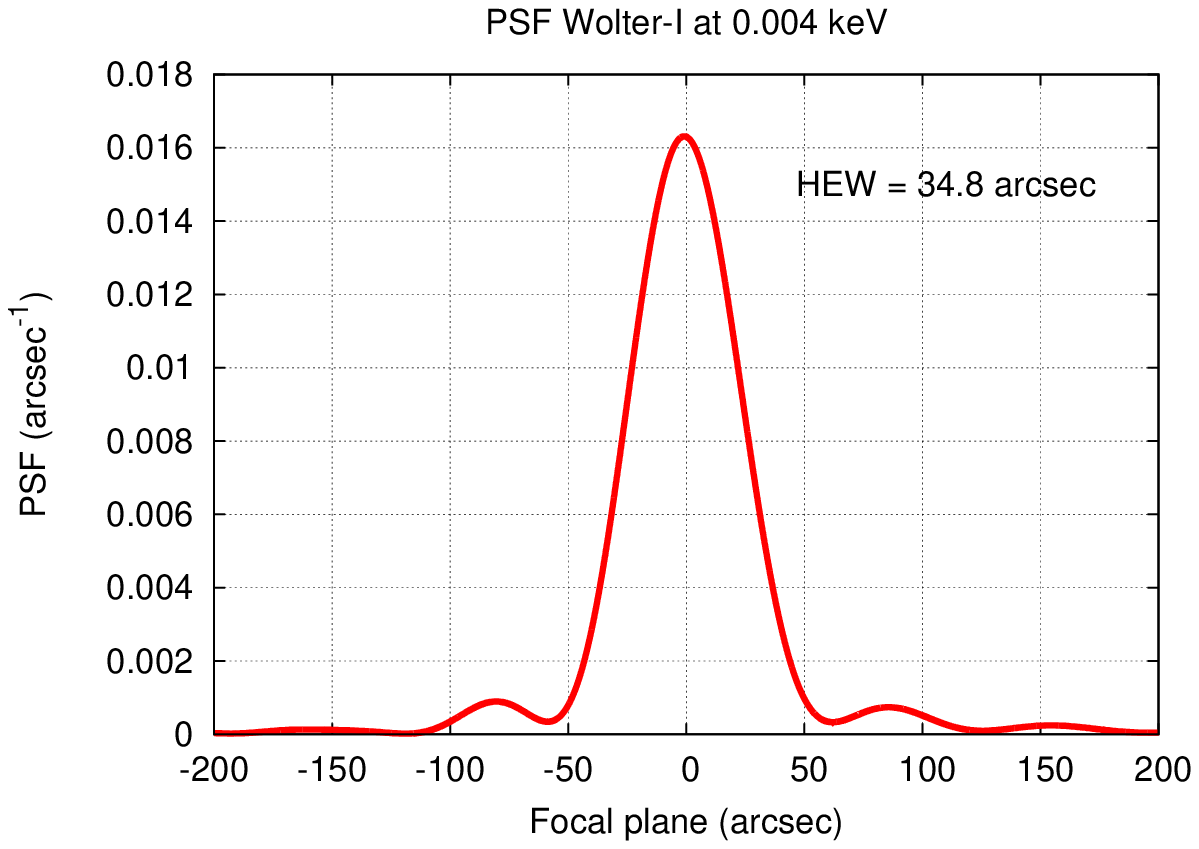}\label{fig:WI_PSF_1a}}
     \subfigure[]{\includegraphics[width=0.45\textwidth]{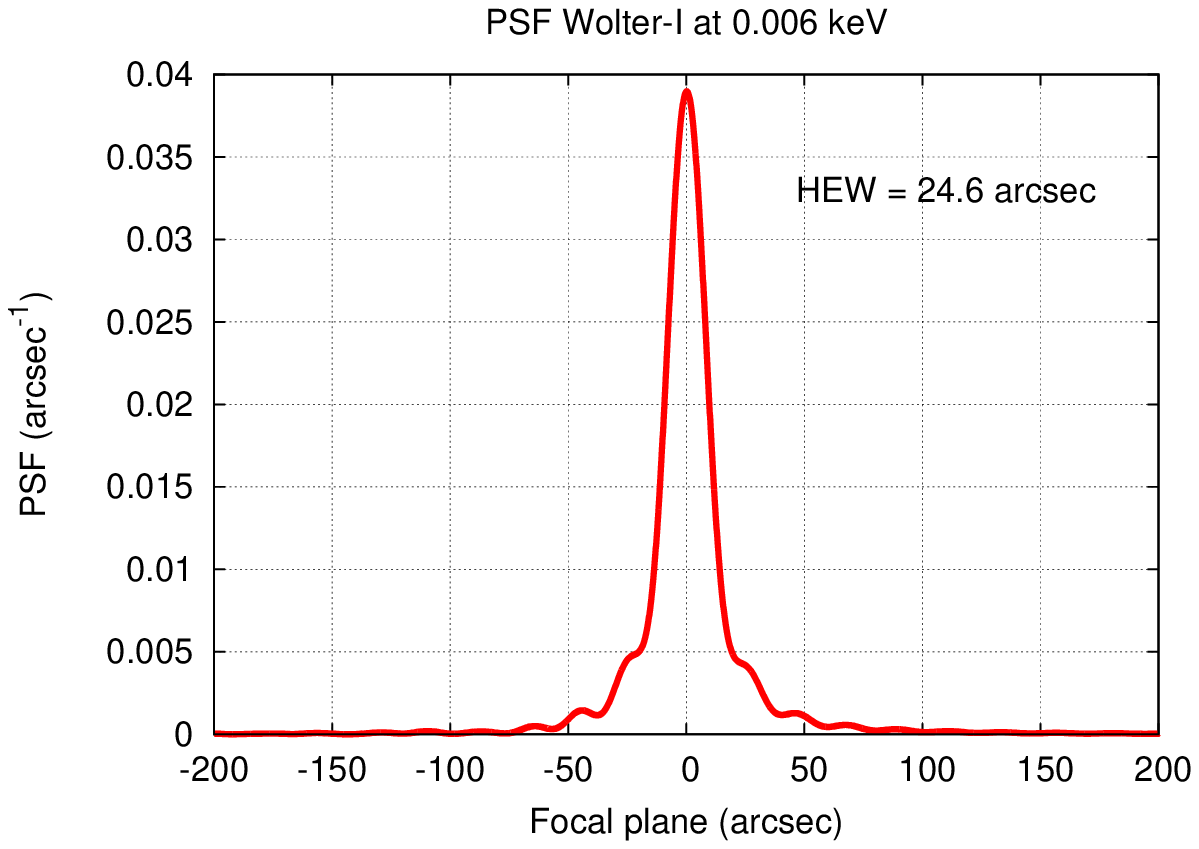}\label{fig:WI_PSF_1b}}
    \subfigure[]{\includegraphics[width=0.45\textwidth]{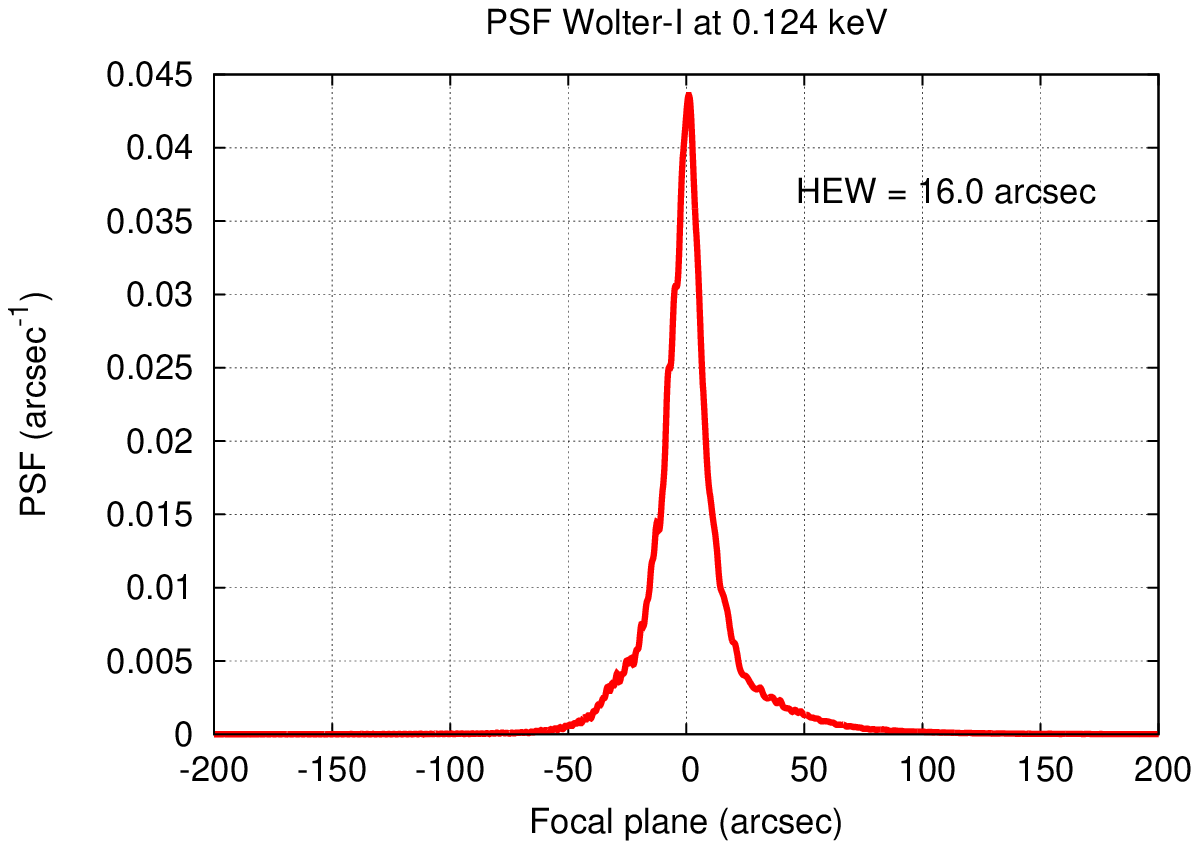}\label{fig:WI_PSF_1c}}
    \subfigure[]{\includegraphics[width=0.45\textwidth]{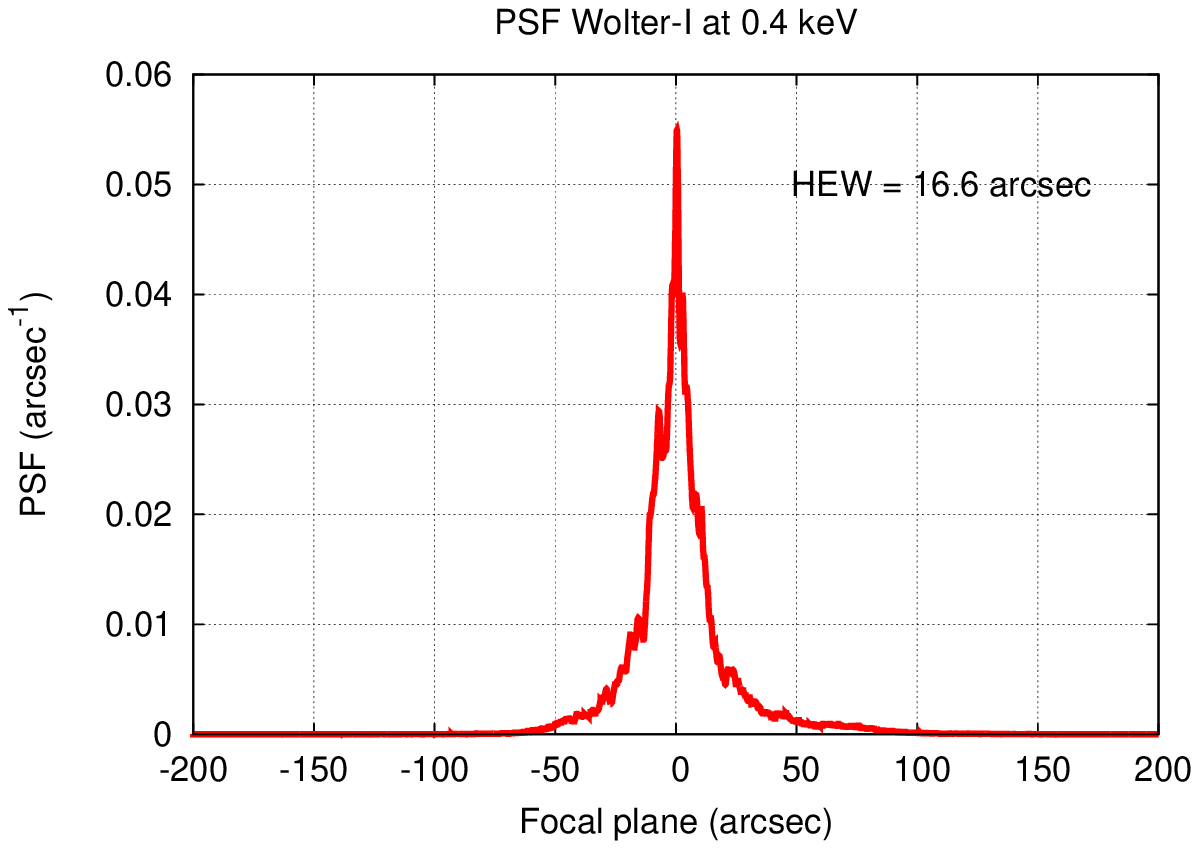}\label{fig:WI_PSF_1d}}
    \caption{WISE results from near-UV to soft X-rays for a Wolter-I mirror including a long-period error and roughness. (a) $\lambda$ = 3000~\AA: the aperture diffraction conceals most of the mirror defects, the HEW is the same as that a perfect mirror of the same size (Fig.~\ref{fig:perfect_mirror_a}). (b) $\lambda$ = 2000~\AA: the aperture diffraction is reduced and the PSF due to mirror shape becomes visible. (c) $\lambda$ =100~\AA; aperture diffraction fringes have completely disappeared, and the PSF is almost equal to the ray-tracing result from the the sole figure. Like the PSF in Fig.~\ref{fig:powlaw}, the result is slightly asymmetrical on the side of the negative angle. (d) $\lambda$ = 30~\AA. The PSF is still dominated by the figure, but some noisy features and a slight HEW increase already announce the appearance of the scattering.}
    \label{fig:PSF_UV2X}
\end{figure*}

For simplicity, we assume the PSD to be expressed by a power-law function of the spatial frequency $\nu$ (Church~\cite{Church88}),
\begin{equation} 
	P(\nu) = \frac{K_n}{\nu^{n}},
	\label{eq:powerlaw}
\end{equation}
where the spectral index $n$ ($1<n<3$) and the coefficient $K_n$ depend on the surface finishing level. The rough profiles of lengths $L_1$ and $L_2$, and resolutions $\Delta z_1$ and $\Delta z_2$ (Eqs.~(\ref{eq:minsamp_z1}) and~(\ref{eq:minsamp_z2})), are generated from Eq.~(\ref{eq:powerlaw}) adopting the parameter values $n$ = 1.5, $K_n$~= 150~nm$^3~\mu$m$^{-1.5}$. The $k^{th}$ Fourier coefficient amplitude of the $j^{th}$ mirror segment, $|a_{k,j}|$, is computed as $|a_{k,j}| = \sqrt{P(\nu_k)/2L_j}$, where $\nu_k = \pm k/L_j$, $k =1, 2,\ldots, L_j/2\Delta z_j$, and $|a_{0,j}| =0$. The phases of the harmonics are selected at random, with the condition $a_{-k,j} = a_{k,j}^*$ to cancel the imaginary part of the rough profile, which is finally obtained by inverse Fourier transform of the $\{a_{k,j}\}$. Every different choice of the phases returns a different rough profile, but always with the same PSD. 

We have thereby computed the expected PSF at $\lambda = 6~\AA$, corresponding to X-rays of 2~keV, in single (Eq.~(\ref{eq:PSF})) and double reflection (Eqs.~(\ref{eq:field2}) and~(\ref{eq:PSF_W1})). To reduce the PSF noise resulting from the pseudo-random nature of roughness, we repeated the computation with five different pairs of profiles with the same PSD, and averaged the results. The resulting PSFs are plotted in Fig.~\ref{fig:powlaw}, showing a sharp peak with the characteristic skirt of X-ray scattering, characterized by a slight asymmetry. As expected, the double reflection enhances the amount of scattering.

We compared the single-reflection PSF with the scattering diagram -- that is, the normalized scattered power per scattering angle unit -- expected from the classical scattering theory (Church et al.~\cite{Church79}; Stover~\cite{Stover95}), neglecting the polarization factor
\begin{equation} 
	\mbox{PSF}(\theta_{\mathrm s}) = \frac{16\pi^2}{\lambda^3}\sin\alpha_0\sin^2\theta_{\mathrm s}\,\frac{1}{2}P\left(\frac{\cos\alpha_0-\cos\theta_{\mathrm s}}{\lambda}\right),
	\label{eq:XRS_1st}
\end{equation}
where $\theta_{\mathrm s} = \alpha_0 +\theta$ is the scattering angle measured from the surface, and the factor 1/2 in front of the PSD accounts for the scattering parted at positive or negative $\theta$. Equation~(\ref{eq:XRS_1st}) is valid at shallow $\alpha_0$ and for roughness values within the smooth surface limit (Eq.~(\ref{eq:smooth})); that is, for scattering angles not too near the specular direction, otherwise the $P(\nu)$ function diverges to inacceptably high values. With the selected values for the parameters $n$ and $K_n$, we have for the reconstructed profiles $\sigma \approx 10~\AA$. The computed smooth surface limit for the adopted values of $\lambda$ and $\alpha_0$ corresponds to 125~\AA; therefore, the first-order XRS theory can be used. The application of Eq.~(\ref{eq:XRS_1st}) returns a PSF perfectly superposed on the single-reflection findings of the Fresnel diffraction applied to the reconstructed rough profiles (Fig.~\ref{fig:powlaw}). The agreement is perfect for $|\theta| >$ 1~arcsec. For lower values of $|\theta|$, the first-order theory diverges to infinity, while the Fresnel theory remains finite. For the rough Wolter-I mirror, the scattering theory does not provide practical formulae like Eq.~(\ref{eq:XRS_1st}) to simulate a double scattering, while the application of the Fresnel diffraction easily returns a scattering diagram that matches twice the single scattering result very well. This can be interpreted as a superposition of two identical scattering diagrams from the parabola and hyperbola. In other words, the multiple scattering is negligible if the surface is within the smooth surface condition, as De~Korte et~al. already pointed out in 1981.
\begin{figure*}[!tpb]
	\centering
    \subfigure[]{\includegraphics[width=0.45\textwidth]{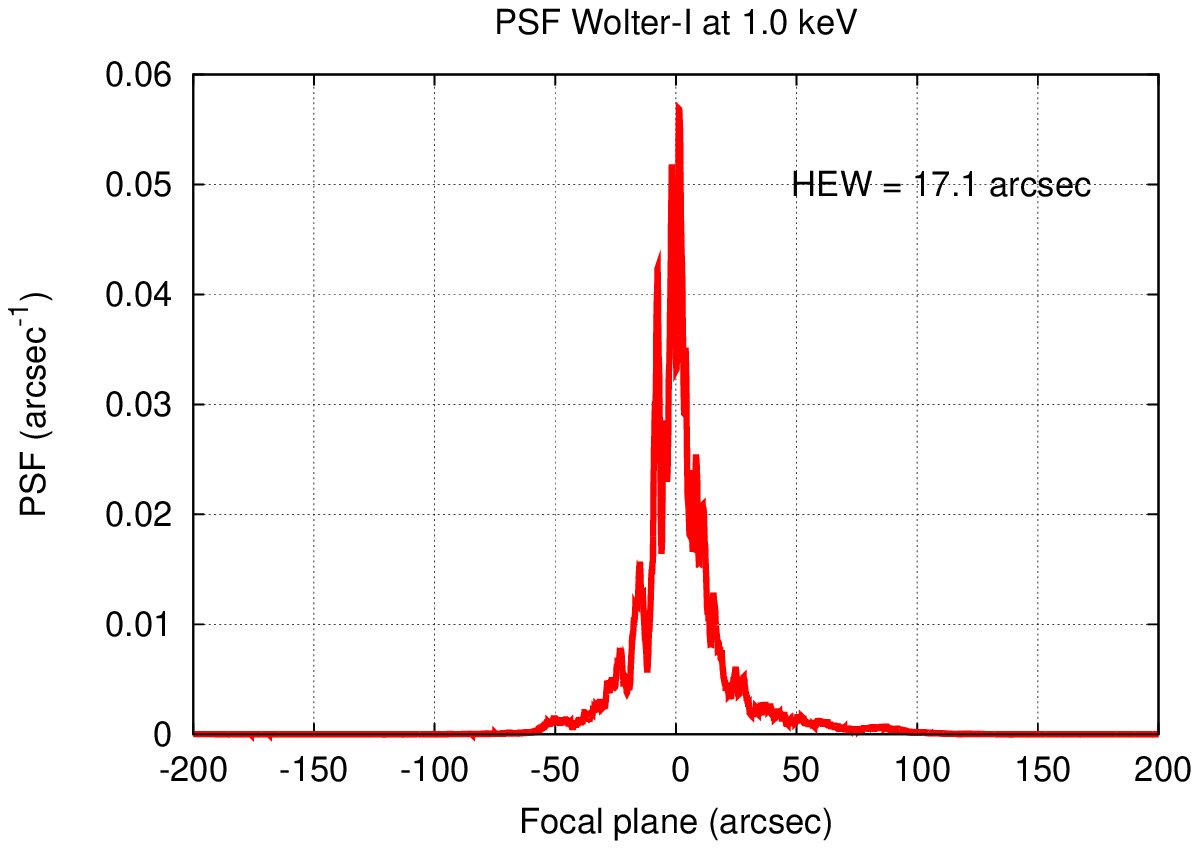}\label{fig:WI_PSF_2a}}
    \subfigure[]{\includegraphics[width=0.45\textwidth]{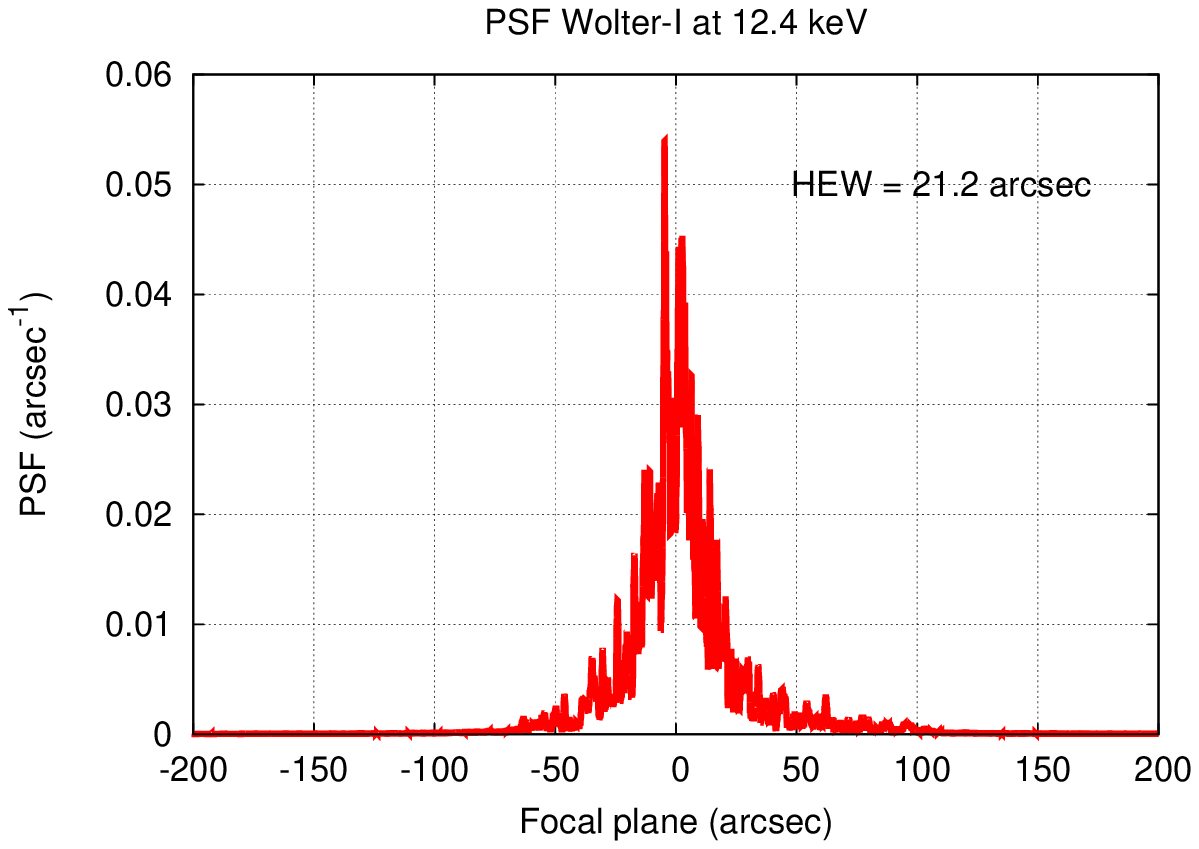}\label{fig:WI_PSF_2b}}
    \subfigure[]{\includegraphics[width=0.45\textwidth]{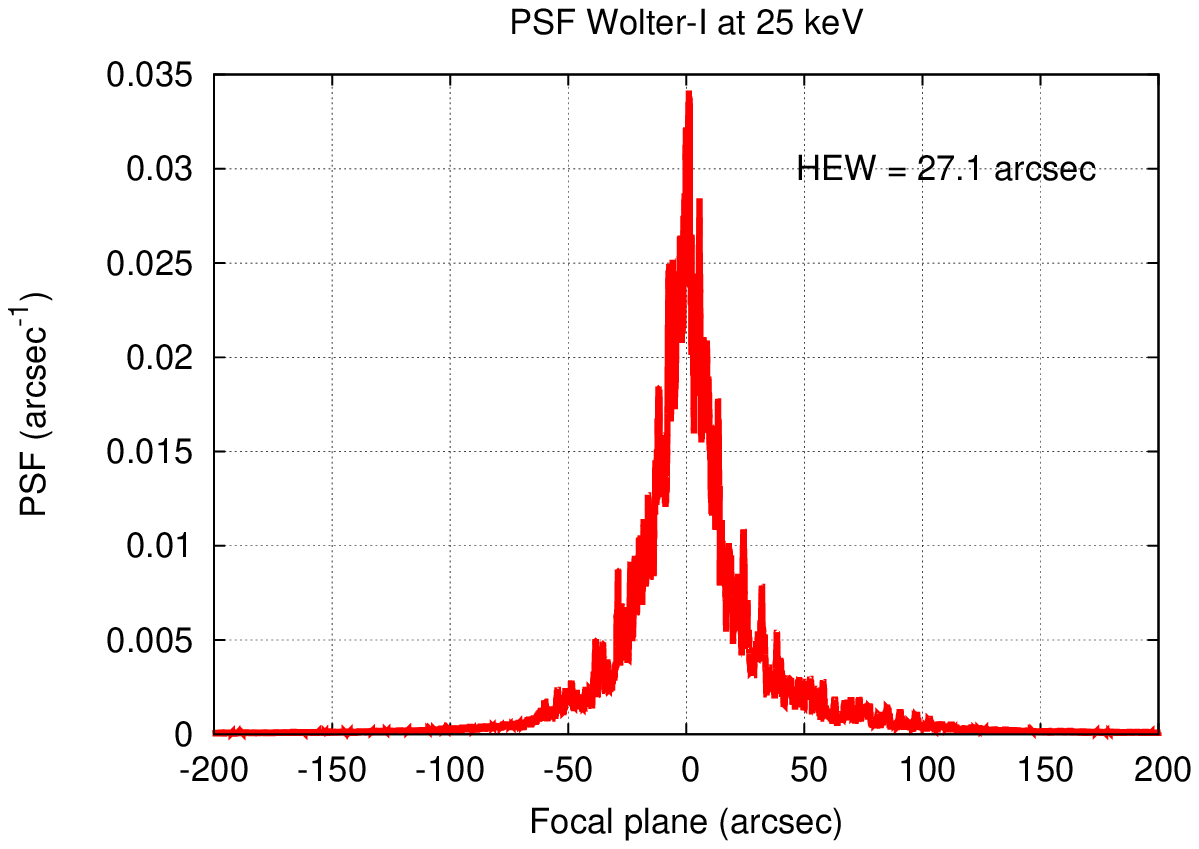}\label{fig:WI_PSF_2c}}
    \subfigure[]{\includegraphics[width=0.45\textwidth]{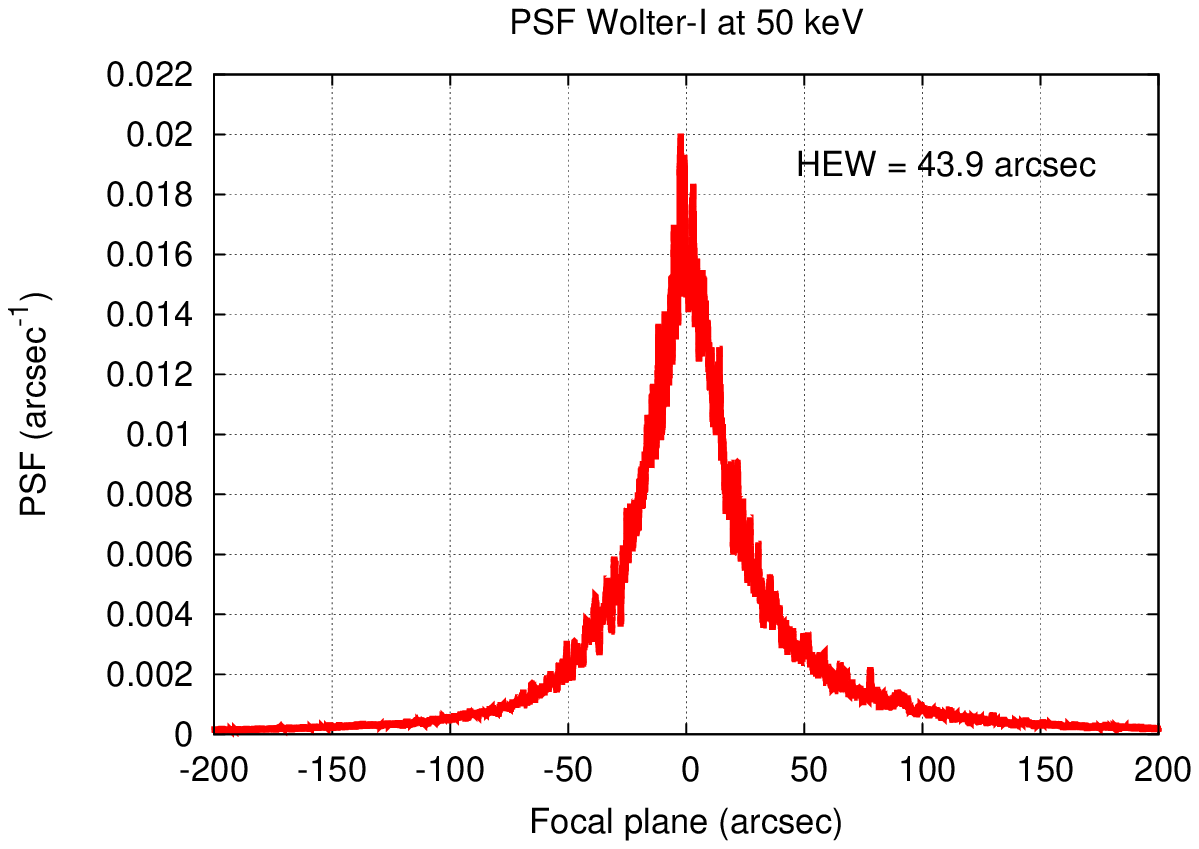}\label{fig:WI_PSF_2d}}
\vspace{-3mm}\caption{WISE results in X-rays: (a) $\lambda$ =12~\AA; the first roughness effects start to appear. (b) $\lambda$ = 1~\AA; the roughness effects are now clearly visible. The PSF is much broader and the HEW increases rapidly. (c) $\lambda$ = 0.5~\AA. (d) $\lambda$ = 0.25~\AA; the X-ray scattering is now overwhelmingly dominant.}
    \label{fig:PSF_X}
\end{figure*}

It is interesting to note that the slight PSF asymmetry, which stems from the $\sin^2\theta_s$ factor appearing in Eq.~(\ref{eq:XRS_1st}), is also reproduced accurately. We conclude that in the limit of smooth surfaces, the formalism provided here correctly reduces to the known scattering theory and is also able to satisfactorily reproduce the scattering in multiple-reflection systems.

\subsection{Parabola and hyperbola with long-period deformations and roughness}\label{figrough}
In the previous sections we have seen that Eqs.~(\ref{eq:field2}) and~(\ref{eq:PSF_W1}) found in the framework of the Fresnel diffraction are able to accurately reproduce the individual factors that classically degrade the PSF of a Wolter-I mirror: the aperture diffraction (Sect.~\ref{perf}), mid-frequencies (Sect.~\ref{grating}), geometrical errors (Sect.~\ref{longper}), and microroughness (Sect.~\ref{rough}). In all these cases, exactly the same treatment was used, changing only the value of $\lambda$. However, these aspects were hitherto analyzed separately. In this section, we provide a more realistic example of a mirror profile including defects over more than one spectral regime. We adopted the mirror dimensions as in Fig.~\ref{fig:field_hyp1} and as figure error the same profile (Eq.~(\ref{eq:shape}) with $w_1 = w_2$ = 8~arcsec, Fig.~\ref{fig:double_curvature}). The roughness PSD is described by Eq.~(\ref{eq:powerlaw}). We have assumed as realistic parameter values $n$~= 1.8, $K_n$~= 2.2~nm$^3$~$\mu$m$^{-1.8}$ and generated two of the infinitely possible profiles from this power spectrum, as described in Sect.~\ref{rough}, with spatial resolutions determined by Eq.~(\ref{eq:minsamp_z1}) for the parabola and by Eq.~(\ref{eq:minsamp_z2}) for the hyperbola. 

After superposing the rough profiles on the modeled figure errors, the PSF is computed at several wavelengths from ultraviolet light to hard X-rays, always applying the same equations at different values of $\lambda$, that is, Eqs.~(\ref{eq:field2}) and~(\ref{eq:PSF_W1}). As in the previous section, we always took the average of five consecutive simulations to reduce the noise in the PSF. This average is needed only in hard X-rays, however, where the roughness effect starts to become apparent. Finally, the PSF was degraded to a realistic spatial resolution of the detector (20~$\mu$m). 
\begin{figure*}[!htpb]
    \subfigure[]{\includegraphics[width=0.5\textwidth]{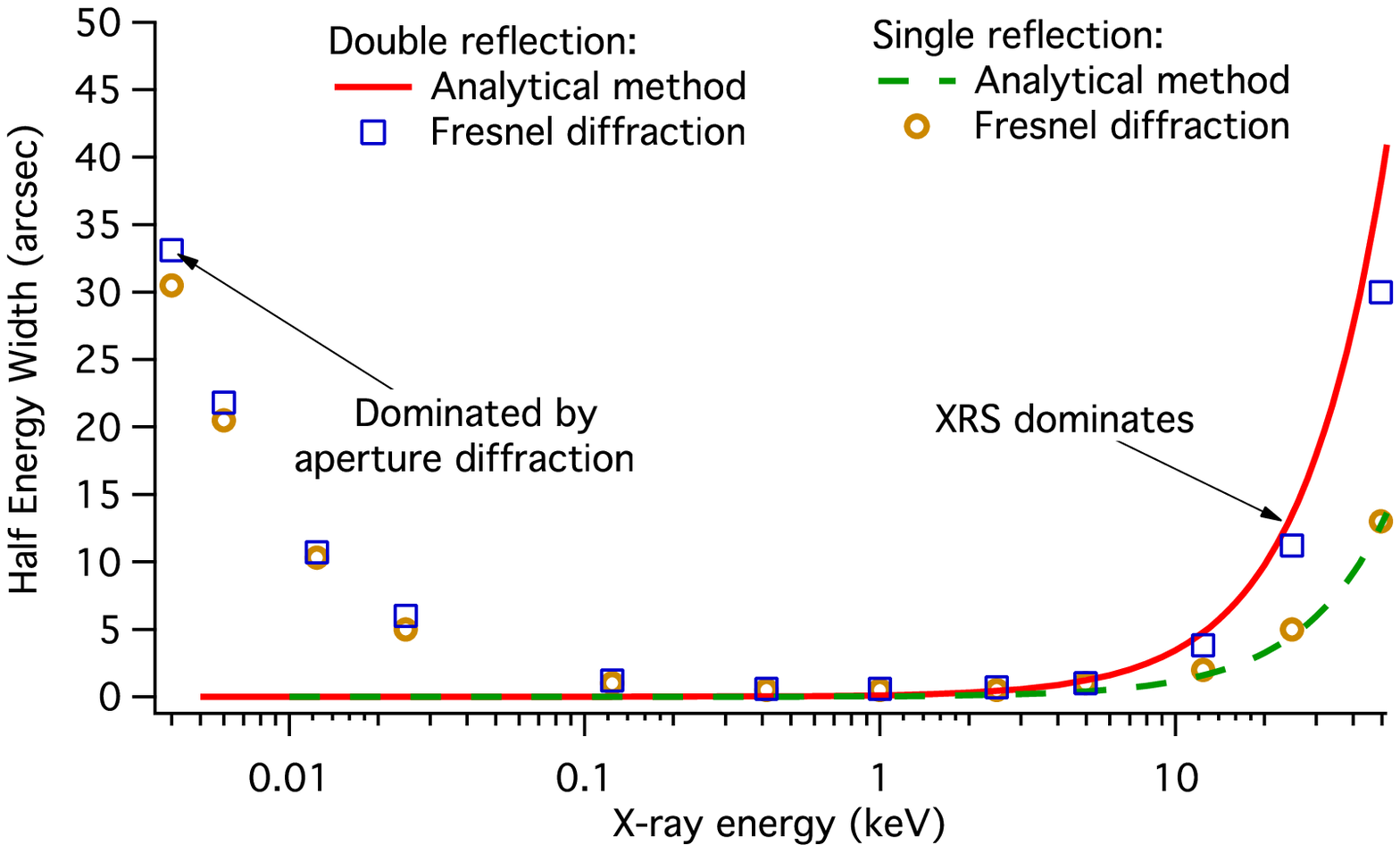} \label{fig:HEW_PSD}}
    \subfigure[]{\includegraphics[width=0.5\textwidth]{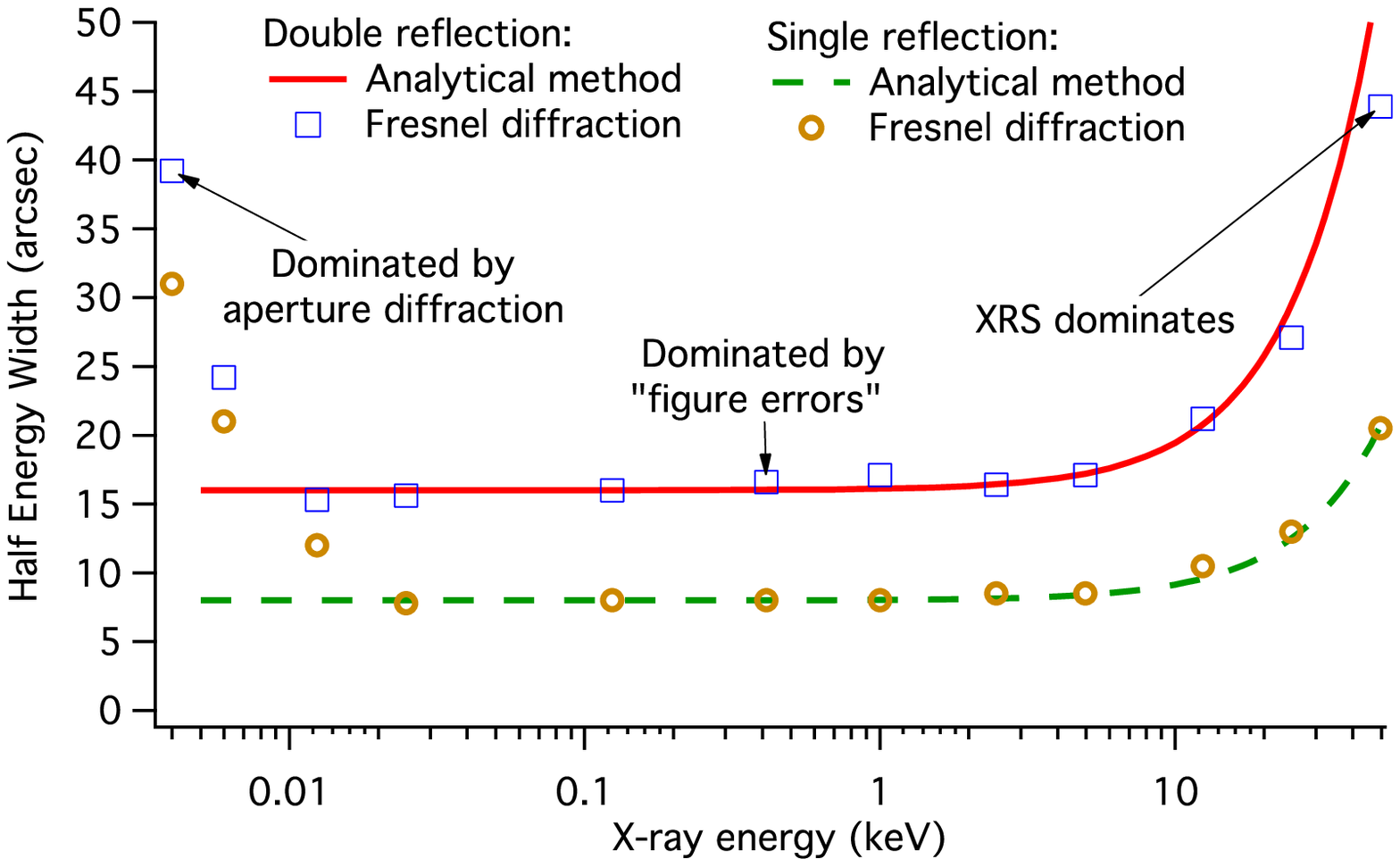} \label{fig:HEW_total}}
    \caption{HEW results as computed by the analytical method (Spiga~\cite{Spiga2007}, lines) and using the WISE code (symbols) for single and double reflection. (a) sole roughness simulated from PSD, (b) roughness and deterministic error. The mirror parameters are the same as adopted in Sect.~\ref{figrough} (after Raimondi \& Spiga~\cite{RaiSpi2011}).}
    \label{fig:HEW_behavior}
\end{figure*}

The calculated PSFs for $\lambda$ ranging from near UV to soft X-rays are reported in Fig.~\ref{fig:PSF_UV2X}. In the UV range, the aperture diffraction pattern is dominant (Fig.~\ref{fig:WI_PSF_1a}) and almost indistinguishable from the pattern of a perfect mirror (Fig.~\ref{fig:perfect_mirror_a}). As $\lambda$ is decreased, the HEW value also decreases but -- unlike in Sect.~\ref{perf} -- does not tend to zero: the aperture diffraction gradually disappears and the mirror deformation effects become visible. In soft X-rays (Fig.~\ref{fig:WI_PSF_1c}), the PSF is almost indistinguishable from the ray-tracing findings as in Sect.~\ref{longper}, and the HEW at $\lambda$ = 100~\AA~equals the 16~arcsec predicted by geometrical optics. This means that the effect of roughness is, in the present case, completely negligible at energies below 0.12~keV.

For $\lambda <$~30~\AA, however, the roughness effect begins to be visible (Fig.~\ref{fig:WI_PSF_1d}). The resulting X-ray scattering causes the PSF to broaden and the HEW to increase in consequence. The scattering effect increases faster and faster as the energy is increased (Fig.~\ref{fig:PSF_X}), gradually concealing the shape of the PSF determined by the deterministic deformation. At 50~keV, the PSF is completely dominated by the scattering (Fig.~\ref{fig:WI_PSF_2d}). 

This example shows that a PSF can be computed in a wide range of $\lambda$ values from a realistic profile, including analytical or measured deformations and microroughness. As anticipated in Sect.~\ref{approx}, we were not required to set boundaries between spectral regimes to treat with different methodologies, and consequently, a combination of the PSFs obtained in the respective spatial frequency ranges was not necessary; the relative weights of geometry and scattering are automatically included in the computed PSF, at each selected value of $\lambda$.

\section{Result validation with the analytical HEW dependence on $\lambda$}\label{HEW}
It is also interesting to investigate the PSF evolution with the energy observing the dependence of its HEW (Sect.~\ref{intro}) on $\lambda$. The computation reported in the last section was performed for the single and the double reflection, and the HEW values obtained are plotted as a function of the energy $E \propto 1/\lambda$ in Fig.~\ref{fig:HEW_total}. To isolate the contribution of the surface roughness to the HEW, the WISE code was also applied to the sole rough profile, without any additional figure error: the results are shown in Fig.~\ref{fig:HEW_PSD} as symbols in the graph. As we anticipated in the previous section, the scattering caused by the surface roughness starts to increase the HEW in X-rays beyond a few keV, and becomes relevant above 10~keV. At low energies, the HEW increase related to the aperture diffraction is clearly visible.

These HEW trends can be validated by comparing them with the analytical method (Spiga~\cite{Spiga2007}) based on the first-order scattering theory to predict the X-ray scattering term of the HEW as a function of $\lambda$, H($\lambda$), given the mirror roughness PSD (and vice versa). The power-law PSD adopted in Sect.~\ref{figrough} fulfills the smooth surface limit (Eq.~(\ref{eq:smooth})) for almost all the energies considered in the computation ($<$~55~keV), therefore the XRS theory at the first order can be applied to a good approximation. If the PSD can be approximated by the Eq.~(\ref{eq:powerlaw}), $H(\lambda)$ can be written in an explicit form:      
\begin{equation}
	H(\lambda) = 2\left[\frac{16\pi^2K_n}{(n-1)\ln \xi}\right]^{\frac{1}{n-1}}\left(\frac{\sin\alpha_0}{\lambda}\right)^{\frac{3-n}{n-1}},
\label{eq:HEW_powerlaw}
\end{equation}
where $\xi$~=~2 for the single reflection and $\xi$~=~4/3 for the double reflection (Wolter-I).

The $H(\lambda)$ trends computed from Eq.~(\ref{eq:HEW_powerlaw}) for the parameter values $n$~= 1.8, $K_n$~= 2.2~nm$^3$~$\mu$m$^{-1.8}$ are reported as lines in Fig.~\ref{fig:HEW_PSD}. At low energies, the trends differ because the Fresnel diffraction method also accounts for aperture diffraction, while Eq.~(\ref{eq:HEW_powerlaw}) does not. At high energies, the HEW computed with the two methods increase in mutual accord: the slight HEW overestimation with the analytical formula at higher energies can be due to the small scattering angles approximation, required by the first-order XRS theory, which is not exactly fulfilled, while the Fresnel diffraction method does not require this condition. The low- and high-energy regimes are separated by a wide plateau where neither the aperture diffraction nor the scattering are relevant. Since no profile error other than roughness is assumed in Fig.~\ref{fig:HEW_PSD}, the HEW plateau is close to zero.

The HEW trend in Fig.~\ref{fig:HEW_total} obtained by the WISE code appears to be similar to the trend in Fig.~\ref{fig:HEW_PSD}, but the mid-energy plateau is at a 8 arcsec for the single reflection and at a 16 arcsec for the Wolter-I. These figure error HEW values are the same as were obtained from the computation in Sect.~\ref{longper} using the sole profile errors, taken with the same sign, at an X-ray energy where the PSF predictions merge with the ray-tracing result. The analytical simulations match the Fresnel diffraction results for the single and double reflection only if the respective figure error HEW values are added linearly to the $H(\lambda)$ functions computed from the sole PSD. In other words, if the XRS and the figure error terms of the HEW can be computed separately, they are to be combined linearly (Raimondi \& Spiga~\cite{RaiSpi2010, RaiSpi2011}) and not quadratically, as initially assumed. Hence, this example not only provides a crossed validation of the analytical and the Fresnel approaches, but also suggests the correct way to mix the two contributions. 

We recall that this comparison was possible because there were no relevant mid-frequency deformations, which would have hindered not only the application of the scattering theory, but also the isolation of an energy regime where the profile error could have been treated according to geometrical optics.

\section{Experimental verification}\label{exper}
The results obtained by our method were also been validated experimentally. A replicated Wolter-I mirror shell in Nickel-Cobalt alloy with a W/Si multilayer coating was manufactured by Media-Lario Technologies (MLT, Bosisio Parini, Italy) and INAF/OAB as a demonstrator for the optics of the hard X-ray NHXM telescope project (now cancelled). The mirror shell was initially measured in full illumination at the PANTER facility (Burwitz~et al.~\cite{Burwitz}) at 1 to 40~keV. Subsequently, the PSF at 15 to 63~keV was measured sector-by-sector at the beamline BL20B2 of the SPring-8 radiation facility (Ogasaka~et al.~\cite{Ogasaka}). The mirror shell with the details of the manufacturing process is described in Basso~et al.~\cite{Basso2011}. Details on the experimental setup and the results obtained at SPring-8 are given in Spiga~et al.~\cite{Spiga2011}. Since the HEW at 1 keV (measured at PANTER) was 18 arcsec and the HEW values measured at SPring-8 are much higher, the PSF at energies above 20 keV are probably already sensing the effect of the surface roughness. Aiming at quantitatively explaining this effect, we compared the experimental results with the PSF simulations following the method described in this paper, accounting for the metrology of profile and roughness.

After the direct measurement in X-rays, the mirror shell longitudinal profile was measured using the dedicated mirror shell profilometer at MLT (Sironi et al.~\cite{SironiSPR}), taking six mirror profiles in as many sectors of the integrated mirror shell, with a 0.4~mm resolution. Finally, the shell was dismounted, cut into pieces and its roughness PSD was characterized at INAF/OAB using an AFM to cover the spectral region from 50~$\mu$m down to 5~nm of spatial wavelengths. The remaining spectral gap from 1~mm to 50~$\mu$m could not be filled by a direct PSI measurement because of the pronounced azimuthal curvature of the samples. Nevertheless, that spectral gap was covered assuming the roughness PSD of the mandrel, which had been measured before the shell replication. This choice was based on previous experience (Sironi~et al.~\cite{Sironi}) that proved that the roughness mandrel in this spectral band is completely copied by the replicated shell. The entire characterization is also reported in Spiga~et al.~\cite{Spiga2011}.

\begin{figure}[!htpb]
	\centering
	\subfigure[]{\includegraphics[width=0.5\textwidth]{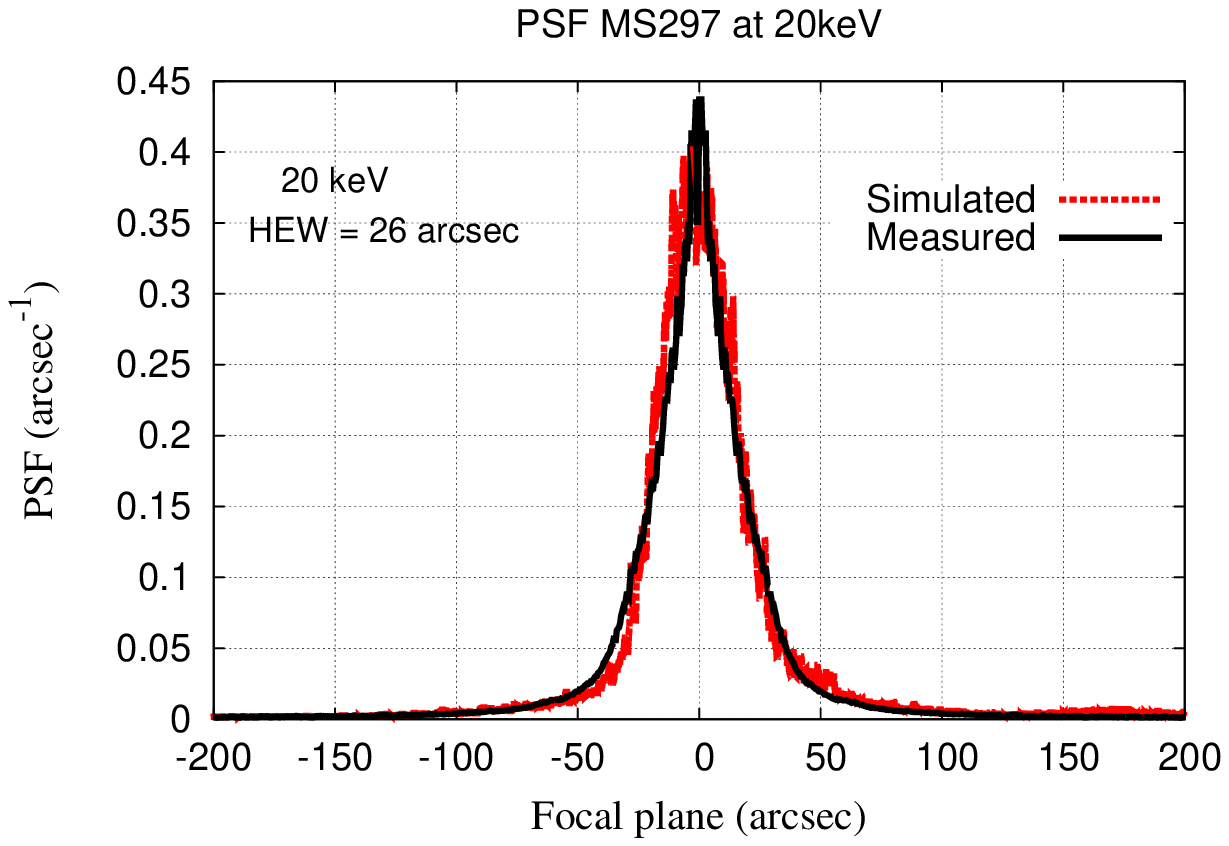} \label{fig:20keV}}
	\subfigure[]{\includegraphics[width=0.5\textwidth]{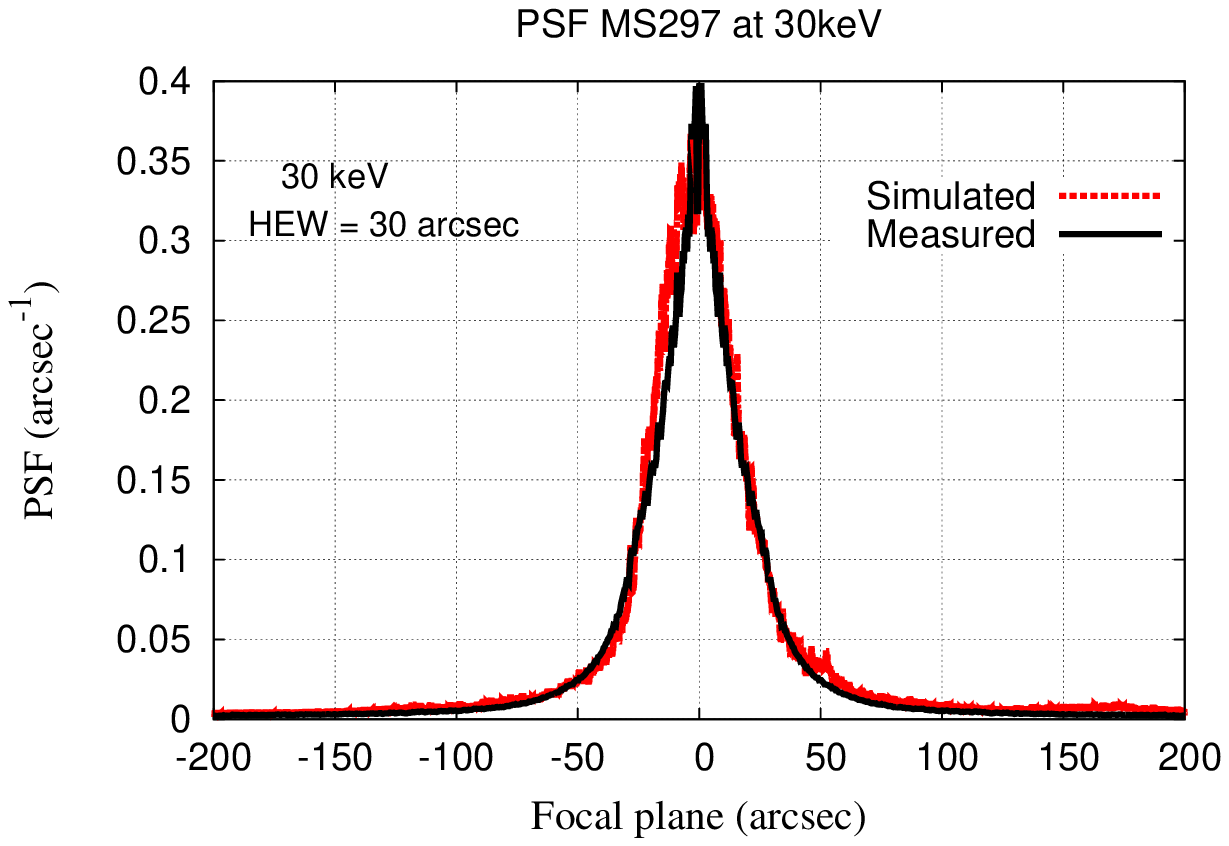} \label{fig:30keV}}
	\subfigure[]{\includegraphics[width=0.5\textwidth]{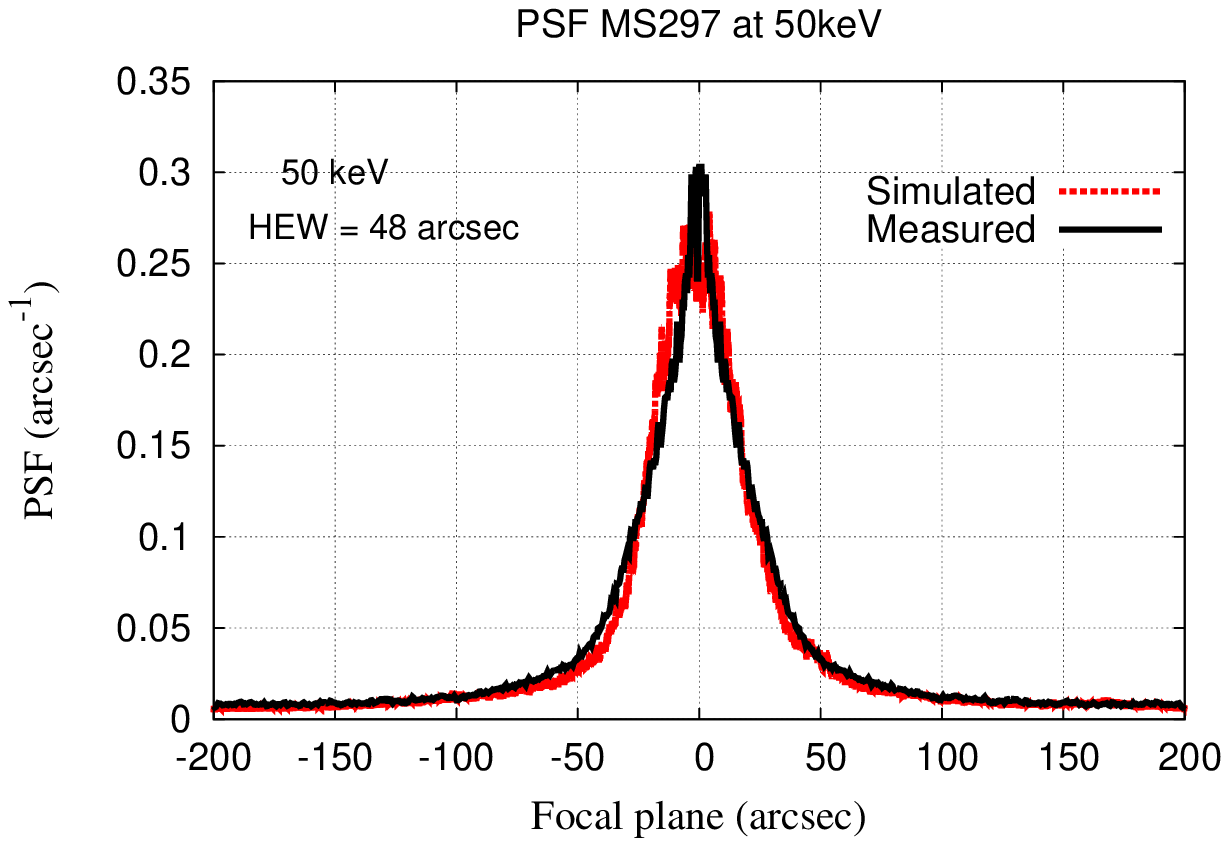} \label{fig:50keV}}
	\caption{Simulated PSF from measured profiles and roughness at 20, 30, and 50 keV using the WISE code to implement the Fresnel diffraction approach. The experimental PSFs are reproduced accurately (after Spiga~et al.~\cite{Spiga2011}).}
	\label{fig:PSF}
\end{figure}

The simulations of the PSFs at 20 keV, 30 keV, and 50 keV were obtained by superposing a rough profile, simulated from the PSD, on each one of the six measured longitudinal profiles, and applying Eqs.~(\ref{eq:field2}) and~(\ref{eq:PSF_W1}) implemented in the WISE code. For each considered X-ray energy, the six PSFs were averaged and normalized over a 4 cm wide region, the same lateral size of the detector as was used to reconstruct the PSFs at SPring-8. In Fig.~\ref{fig:PSF} we display the simulated PSFs, compared with the PSFs obtained from the measurements. The matching is accurate. Consequently, the HEW values computed from the simulated PSFs also agree well with the measured values to within a couple of arcseconds. The remaining discrepancies might be caused by the small number of profiles characterized on the mirror shell, which might not be completely representative of the tests performed at SPring-8, in which almost all the mirror surface was probed. Moreover, the experimental PSF results from a focal spot integration over $2\pi$, hence it exhibits an intrinsic symmetry that cannot be exactly reproduced in the simulation. The agreement remains to be very good.

\section{Conclusions}\label{concl}
We have shown how a computation entirely based on the Huygens-Fresnel principle allows us a comprehensive simulation of a grazing-incidence mirror PSF, without the need of devoting separate physical treatments to different ranges of spatial frequencies. Even if wavefront propagation codes are widespread in X-ray optics, they are rarely applied to real mirror surfaces, which are characterized by defects down to spatial wavelengths in the range of a micron or even less. In fact, the computation would require a double integration over a huge data matrix representing the mirror surface, and in addition, for every point of the detector. In contrast, reducing the formalism to a single dimension -- that is, in the incidence plane -- simplifies the computation by several orders of magnitude. The price to pay is that transverse deviations caused by profile errors in the azimuthal directions are neglected: this effect is usually negligible in practical cases, since their impact is smaller by orders of magnitude than the errors in the longitudinal direction. The 1D Fresnel integral can be numerically computed at any value of $\lambda$, automatically assigning the correct weight to aperture diffraction, geometry, and scattering. The PSF computation \`{a} la Fresnel is self-consistent and versatile: it works in near- and far-field condition, regardless of the finiteness of the source distance, and can be easily extended to multiple reflections as in the case of the Wolter-I design, which is frequently adopted in mirrors for X-ray telescopes. Finally, the formulae can be easily implemented in any computer language: in particular, our WISE code in IDL language yielded results that agreed well with the experiments performed in hard X-rays. Future developments will be aimed at extending the results to 2D images, but still avoid becoming too demanding from a computational point of view.

\begin{acknowledgements}
C.~Brizzolari (Universit\`a di Milano-Bicocca), C.~Pelliciari (INAF/OAB), B.~Salmaso (INAF/OAB and Universit\`a dell'Insubria), and K.~Tayabaly (Politecnico di Milano-Bovisa) are acknowledged for useful discussions. 
\end{acknowledgements}

\appendix
\section{Deriving the integral formula for the diffracted field}
\label{derivation}
In this appendix we compute the electric field in detail (Eq.~(\ref{eq:field})), in scalar approximation, diffracted in the $xz$ plane by a grazing incidence focusing mirror sector (refer to Fig.~\ref{fig:refframe}) with axial symmetry around the $z$-axis and angular aperture $\Delta \Phi$. We assume the focal plane to be at $z =0$, the mirror exit pupil at $z = f$, and the entrance pupil at $f+L_1$, where $L_1$ is the length of the mirror along $z$. The radial coordinate of the mirror is described by the generic function $r_1(z_1)$ assumed to be equal to the profile $x_1(z_1)$ in the $xz$ plane, and varying from $R_0$ to $R_{\mathrm M}$ as $f < z_1 < f+L_1$, with $R_0 \ll f$ and $L_1 \ll f$. Because the roundness error effect is assumed to be negligible owing to the shallow incidence angles (Sect.~\ref{approx}), the mirror surface is described by the coordinates ($r_1 \cos\varphi_1$, $r_1 \sin\varphi_1$, $z_1$). 

A spherical wave propagates in the negative $z$ direction either from (diverging wave) or toward (converging wave) a point located on the $z$-axis at $z = S$. In the first case $S \gg 0$, in the second case $S \ll 0$, and we set $D = S-f$. To simplify the notation, we arbitrarily take as a phase reference the mirror's entrance pupil plane, and temporarily assume the incident electric field amplitude to be a constant, $E_0$. We also denote with $d_1(z_1)$ the distance from the reference wavefront to a point on the mirror, and with $d_2(\varphi_1, z_1, x, z)$ the distance from that point to ($x$, 0, $z$). Finally, be $\delta = r_1/D$ the beam divergence angle at the mirror, in practice constant ($\delta \approx R_0/D$) throughout the mirror surface owing to the high value of $|D|$. Taking $\delta$ with its sign, the incidence angle on the mirror is also constant to a good approximation: $\alpha_1 = \alpha_0+\delta$, where $\alpha_0$ is the incidence angle for a source at infinity (see also Spiga~et al.~\cite{Spiga2009}). The radial aperture of the mirror seen from the source is $\Delta R_1 = L_1 \sin\alpha_1$. 

Owing to the large distance of the source, its intensity decreases negligibly over the mirror length and the mirror is illuminated uniformly. Neglecting obliquity factors, the well-known expression of the secondary electric field in the $xz$ plane, as generated by the mirror area element at the cylindrical coordinates $(r_1, \varphi_1, z_1)$ is
\begin{equation}
	\mbox{d}^2E(x, z)=\frac{E_0}{\lambda d_2}\exp\left[-\frac{2\pi\mathrm{i}}{\lambda}\left(d_1+d_2 \right) \right]\, \mbox{d}^2\underline{x}_{1\perp},
	\label{eq:HF_eq}
\end{equation}
where $\mbox{d}^2\underline{x}_{1\perp} \!= r_1 \,\mbox{d}\varphi_1\,\mbox{d}x_1$ is the area element perpendicular to the beam direction.

For of a diverging spherical wavefront ($S \gg 0$), the distance $d_1$ can be written as 
\begin{equation}
	d_1= f+L_1-S + \sqrt{r_1^2+(S-z_1)^2},
	\label{eq:d1_spheric}
\end{equation}
where the term $f+L_1-S$ was added because the phase reference was chosen at $z = f+L_1$. Since $r_1 \ll S-z_1$, we can approximate Eq.~(\ref{eq:d1_spheric}) as
\begin{equation}
	d_1\simeq f+L_1-z_1+\frac{r_1^2}{2(S-z_1)}.
	\label{eq:d1_spherical}
\end{equation}
Equation~(\ref{eq:d1_spherical}) is also valid for $S \ll 0$ and reduces -- as expected -- to $d_1\simeq f+L_1-z_1$ for a source at infinity. Since the source is on-axis, $d_1$ is independent of $\varphi_1$. In contrast, the $d_2$ dependence on $\varphi_1$ must be explicitly considered to also enable the diffracted field computation off the $z$-axis:
\begin{equation}
	d_2 = \sqrt{r_1^2+x^2+(z_1-z)^2-2 x\, r_1 \cos\varphi_1}.
	\label{eq:d2_expr}
\end{equation}

To avoid diffraction effects at the azimuthal ends, $R_0\Delta\Phi$ should be taken much larger than $\lambda$. To this end, in practical cases it is sufficient for $\Delta\Phi$ to not exceed a few degrees. Then we can assume $\cos\varphi_1 \simeq 1- \varphi_1^2/2$, which turns Eq.~(\ref{eq:d2_expr}) into 
\begin{equation}
	d_2 \simeq \sqrt{\bar{d}_2^2+ x\, r_1 \varphi_1^2},
	\label{eq:d2_app1}
\end{equation}
where we have defined 
\begin{equation}
	\bar{d}_2 = \sqrt{(r_1-x)^2+(z_1-z)^2}.
	\label{eq:d2_bardef}
\end{equation}
Expanding the square root in series in Eq.~(\ref{eq:d2_app1}), we then obtain
\begin{equation}
	d_2 = \bar{d}_2 +\frac{x \,r_1 \varphi_1^2}{2\bar{d}_2}- \frac{x^2 \,r_1^2 \varphi_1^4}{8\bar{d}^3_2}+\cdots \, :
	\label{eq:d2_app2}
\end{equation}
for $z$ not too close to $f$, we are allowed to neglect the terms in Eq.~(\ref{eq:d2_app2}), from the third term on. We substitute the first two terms into Eq.~(\ref{eq:HF_eq}) and obtain by integrating over the mirror surface 
\begin{equation}
	E(x, z)=\!\! \int_{f}^{f+L_1}\!\!\!\! \mbox{d}z_1\, r_1\int_{-\frac{\Delta\Phi}{2}}^{+\frac{\Delta\Phi}{2}}\!\!\!\!\mbox{d}\varphi_1\,\frac{E_0\sin\alpha_1}{\lambda d_2}\,e^{-\frac{2\pi \mathrm{i}}{\lambda}\left(\bar{d}_2+\frac{x\, r_1 \,\varphi_1^2}{2\bar{d}_2} +d_1\right)},
	\label{eq:HF_eq_bis}
\end{equation}
where we have used the relation $\mbox{d}x_1 \simeq \sin\alpha_1\, \mbox{d}z_1$. Recalling the definition of $\Delta R_1$ and approximating $d_2 \approx \bar{d}_2$ in the denominator, we obtain
\begin{equation}
	E(x, z)= \frac{E_0\, \Delta R_1}{L_1\lambda}\!\int_{f}^{f+L_1}\!\!\!\!\!\!\! \mbox{d}z_1\,\frac{r_1}{\bar{d}_2}\,e^{-\frac{2\pi \mathrm{i}}{\lambda}\left(\bar{d}_2 +d_1\right)}\!\!\int_{-\frac{\Delta\Phi}{2}}^{+\frac{\Delta\Phi}{2}}\!\!\mbox{d}\varphi_1\,\,e^{-\mathrm{i}\frac{\pi x\, r_1}{\bar{d}_2\lambda}\varphi_1^2}
	\label{eq:HF_eq_ter}
\end{equation}
and, by defining the dimensionless parameter
\begin{equation}
	 \zeta = \varphi_1\sqrt{\frac{x \,r_1}{\lambda \bar{d}_2}},
	\label{eq:w_def}
\end{equation}
the integral in $\varphi_1$ takes a simpler form
\begin{equation}
\int_{-\frac{\Delta\Phi}{2}}^{+\frac{\Delta\Phi}{2}}\!e^{-\mathrm{i}\frac{\pi x\, r_1}{\bar{d}_2\lambda}\varphi_1^2} \,\mbox{d}\varphi_1 \stackrel{r_1\Delta\Phi \gg \lambda} {\longrightarrow}\sqrt{\frac{\lambda \bar{d}_2}{x \,r_1}}\int_{-\infty}^{+\infty} e^{-\mathrm{i}\pi \zeta^2} d\zeta,
	\label{eq:az_int}
\end{equation}
where the condition $r_1 \,\Delta\Phi\gg \lambda$ has allowed us to approximate the integration limits with infinity. The integral in $\zeta$ is the well-known Fresnel integral and can easily be computed as $(1-i)/\!\sqrt{2}$. We remain with
\begin{equation}
\int_{-\frac{\Delta\Phi}{2}}^{+\frac{\Delta\Phi}{2}}\!e^{-\mathrm{i}\frac{\pi x\, r_1}{\bar{d}_2\lambda}\varphi_1^2} \,\mbox{d}\varphi_1 = \sqrt{\frac{\lambda \bar{d}_2}{x \,r_1}}e^{-\mathrm{i}\frac{\pi}{4}}.
	\label{eq:az_int_bis}
\end{equation}
Finally, substituting this result into Eq.~(\ref{eq:HF_eq_ter}) and neglecting the constant phase factor yields 
\begin{equation}
	E(x, z)= \frac{E_0\, \Delta R_1}{L_1\sqrt{x\lambda}}\!\int_{f}^{f+L_1}\!\!\!\sqrt{\frac{r_1}{\bar{d}_2}}\,e^{-\frac{2\pi \mathrm{i}}{\lambda}\left(\bar{d}_2 +d_1\right)}\, \mbox{d}z_1,
	\label{eq:HF_fin}
\end{equation}
then substituting $d_1$ into Eq.~(\ref{eq:HF_eq_ter}) with the expression provided by Eq.~(\ref{eq:d1_spherical}), disregarding constant phase factors, and replacing $r_1=x_1(z_1)$, exactly yields Eq.~(\ref{eq:field}) in Sect.~\ref{field}. 

Extending the computation for a mirror sector to a source off-axis is straightforward by changing the definition of $d_1$ in Eq.~(\ref{eq:d1_spherical}) to account for the position $x_{\mathrm s}$ of the source in the $xz$ plane:
\begin{equation}
	d_1\simeq f+L_1-z_1+\frac{(x_{\mathrm s}-r_1)^2}{2(S-z_1)}
	\label{eq:d1_spherical_off}
\end{equation}
for a source at finite distance, and 
\begin{equation}
	d_1\simeq f+L_1-z_1-\theta_{\mathrm s} \,r_1
	\label{eq:d1_planar_off}
\end{equation}
for a source at infinity, denoting with $\theta_{\mathrm s}$ the off-axis angle. The definition of $\Delta R_1$ used in Eq.~(\ref{eq:HF_fin}) also needs to be changed to account for the off-axis angle: $\Delta R_1 = L_1\sin(\alpha_0+\delta-\theta_{\mathrm s})$.

\end{document}